%% file: manuscript.tex
\documentclass[fleqn,usenatbib]{mnras}

\usepackage{newtxtext,newtxmath}

\usepackage[T1]{fontenc}

\DeclareRobustCommand{\VAN}[3]{#2}
\let\VANthebibliography\thebibliography
\def\thebibliography{\DeclareRobustCommand{\VAN}[3]{##3}\VANthebibliography}

\usepackage{graphicx}	
\usepackage{amsmath}	
\usepackage{array}
\usepackage[table,x11names]{xcolor}

\newcommand{\Msun}{\ensuremath{\mathrm{M}_{\odot}}}
\newcommand{\eLz}{$e-L_\mathrm{z}$}
\newcommand{\JRLz}{$\sqrt{J_\mathrm{R}}-L_\mathrm{z}$}

\defcitealias{lane22}{LBM22}
\defcitealias{mackereth20}{MB20}

\title[The stellar mass GS/E]{The stellar mass of the \textit{Gaia}-Sausage/Enceladus accretion remnant}

\author[J. M. M. Lane et al.]{
James M. M. Lane$^{1}$\thanks{E-mail: lane@astro.utoronto.ca},
Jo Bovy$^{1}$, \&
J. Ted Mackereth$^{1,2,3}$
\\
$^{1}$David A. Dunlap Department of Astronomy and Astrophysics, University of Toronto, 50 St. George Street, Toronto ON, M5S 3H4, Canada\\
$^{2}$Canadian Institute for Theoretical Astrophysics, University of Toronto, 60 St. George Street, Toronto ON, M5S 3H8, Canada\\
$^{3}$Dunlap Institute for Astronomy and Astrophysics, University of Toronto, 50 St. George Street, Toronto ON, M5S 3H4, Canada
}

\date{Accepted XXX. Received YYY; in original form ZZZ}

\pubyear{2023}

\begin{document}
\label{firstpage}
\pagerange{\pageref{firstpage}--\pageref{lastpage}}
\maketitle

\begin{abstract}
The \textit{Gaia}-Sausage/Enceladus (GS/E) structure is an accretion remnant which comprises a large fraction of the Milky Way's stellar halo. We study GS/E using high-purity samples of kinematically selected stars from APOGEE DR16 and \textit{Gaia}. Employing a novel framework to account for kinematic selection biases using distribution functions, we fit density profiles to these GS/E samples and measure their masses. We find that GS/E has a shallow density profile in the inner Galaxy, with a break between 15--25~kpc beyond which the profile steepens. We also find that GS/E is triaxial, with axis ratios 1:0.55:0.45 (nearly prolate), and the major axis is oriented about 80~degrees from the Sun--Galactic centre line and 16 degrees above the plane. We measure a stellar mass for GS/E of $1.45\,^{+0.92}_{-0.51}\,\mathrm{(stat.)}\,^{+0.13}_{-0.37} \mathrm{(sys.)}\ \times10^{8}$~\Msun. Our mass estimate is lower than others in the literature, a finding we attribute to the excellent purity of the samples we work with. We also fit a density profile to the entire Milky Way stellar halo, finding a mass in the range of $6.7-8.4 \times 10^{8}$~\Msun, and implying that GS/E could make up as little as 15-25~per~cent of the mass of the Milky Way stellar halo. Our lower stellar mass combined with standard stellar-mass-to-halo mass relations implies that GS/E constituted a minor 1:8 mass-ratio merger at the time of its accretion. 

\end{abstract}




\section{Introduction}

In our $\Lambda$CDM Universe, in which structure is arranged hierarchically, the constituent mass of a galaxy like the Milky Way grows largely due to accretion of smaller nearby structures, which supply gas, stars, and dark matter to the gravitationally dominant host \citep{searle78,white91,helmi99,bullock05}. One of the best records of these accretion processes lies in the distribution and arrangement of stars in the Milky Way halo, which have comparatively dissapationless kinematics and have therefore retained signatures of the events which brought them into the Galaxy \citep{freeman02,johnston08}. Mergers play an outsized role in shaping our Galaxy beyond its stellar and dark halo, and are thought to contribute to the formation or evolution of the bulge, bar, disk, as well as the generation of past and ongoing instabilities and perturbations to these stellar components \citep[][]{bland-hawthorn16,helmi20}. An inventory of the stellar constituents of the Milky Way halo is therefore crucial to understanding the role of mergers in the broader context of the formation and evolution of our Galaxy.

The combination of astrometric data from the \textit{Gaia} mission \citep{gaia} and large ground-based spectroscopic surveys such as SEGUE \citep{segue}, APOGEE \citep{apogee}, GALAH \citep{galah}, LAMOST \citep{lamost}, and H3 \citep{h3}, has proved a boon for studying the Milky Way stellar halo. One of the most intriguing results of this new era is the detailed characterization using data from the second \textit{Gaia} data release of an apparent merger remnant, dubbed \textit{Gaia}-Sausage/Enceladus (GS/E), which dominates the nearby stellar halo \citep{belokurov18,haywood18,helmi18}. While GS/E has been the subject of intense recent study, its synthesis can be traced back before the second \textit{Gaia} data release to the notion of a `dual' galactic halo based on its broken radial density profile and bimodal chemistry \citep[e.g.][]{carollo07,nissen10,deason11,bonaca17,hayes18}. GS/E is characterized by a large group of stars on highly radial orbits, with near-zero angular momentum $L_\mathrm{z}$, a wide span of radial velocities, and high eccentricities. The remnant was quickly seen to comprise a significant fraction ($\approx$ 50~per~cent) of the density of the nearby stellar halo \citep{belokurov18,fattahi19,lancaster19}, and initial estimates of the stellar mass of the progenitor were quite high, ranging from $6\times10^{8}$ to nearly $10^{10}$ \Msun\ \citep{helmi18,belokurov18,vincenzo19,feuillet20,fattahi19,myeong19,das20}. The simplest interpretation of this accretion remnant is that it was deposited during a massive, head-on merger event early in the life of the Milky Way \citep{helmi18,belokurov18,iorio19,deason19,fattahi19,mackereth19a}. Although other works have argued that the observed chemodynamics are also consistent with multiple smaller accretion events \citep{donlon22,donlon23}.

Recent, thorough studies of the GS/E accretion remnant focusing on its density profile have tended to settle on a lower stellar mass for the progenitor in the range $3\times10^{8}$ to nearly $10^{9}$~\Msun\ \citep{mackereth19a,mackereth20,naidu21,han22}. Additionally, the chemical composition of the remnant has received attention, and specifically its [Fe/H] abundance distribution (which extends to high [Fe/H] $\sim -0.6$) as well as its [$\alpha$/Fe] distribution, both of which appear to reflect those of a massive dwarf galaxy \citep{myeong19,monty20,mackereth19a,feuillet20,matsuno21,hasselquist21,buder22,gaiadr3_chemodynamics,horta23a}, supporting the earlier findings of an [Fe/H]-rich structure by \citet{belokurov18} as well as pre-\textit{Gaia} second data release studies \citep{nissen10,bonaca17,hayes18}. The ages of likely GS/E stars were measured by \citet{montalban21} and found to be typically $10$~Gyr, supporting other works which have suggested a similar timeframe for the occurrence of the accretion episode ($z\approx2$). Our understanding of the kinematics of the remnant have been extended by \citet{lancaster19} and \citet{iorio21}, who measure a strong degree of radial anisotropy in the population, behaviour extending over a wide range of Galactocentric radii. \citet{naidu21} and \citet{chandra23} compare H3 data with tailored N-body simulations and develop a coherent narative for the specifics of the infall of the progenitor. They also link several other halo structures such as Arjuna \citep{naidu20}, the Hercules-Aquila cloud, and the Virgo Overdensity with GS/E debris. The association of GS/E with the latter two structures was first proposed by \citet{simion19}, with support coming from the analysis of stellar halo density profiles by \citet{iorio19}, and has recently been affirmed through chemical analysis by \citet{perottoni22}. Given the menagerie of recently discovered halo substructures \citep[see ][ for censuses of the major ones]{helmi20,naidu21,horta23a}, there are open questions regarding which are unique and independent of GS/E, and which are associated with, or simply complex dynamical echoes of, GS/E. 

Recently, \citet[][hereafter LBM22]{lane22} presented kinematic models of the nearby stellar halo based on distribution functions representing the major constituent populations: the metal-rich, radially anisotropic population attributed to GS/E, and the metal poor, comparatively radially isotropic population attributed to the (\textit{in-situ}) remainder of the stellar halo \citep[e.g.][]{belokurov18,haywood18}. \citetalias{lane22} found that kinematic selection criteria commonly used to identify GS/E stars are subject to bias in the context of their models, and that most selection criteria are at best about 70~per~cent, and at worst about 50~per~cent pure (i.e., only 5-7 in every 10 stars attributed to GS/E are genuine members of the GS/E remnant). They also use their models to construct high-purity, scale-free (i.e., they are resilient to changes in stellar sample or gravitational potential used for calculation of kinematic quantities) selections for GS/E, which reach as high as 85~per~cent purity. \citetalias{lane22} highlight that while the GS/E remnant is an overdensity occupying a \textit{characteristic} region of phase-space, it is does not do so without contamination from other stellar populations. They emphasize the need to account for these effects when modelling the GS/E remnant, which is particularly important as studies seek to place tighter constraints on its properties.

In this study we will build on the work of \citet[][hereafter MB20]{mackereth20} who studied the density profile of the stellar halo with data from \textit{Gaia} and the Apache Point Galactic Evolution Experiment \citep[APOGEE][]{apogee}. We integrate the kinematic models of \citetalias{lane22} into the density modelling framework of \citetalias{mackereth20} so that we may select a high-purity sample of GS/E stars based on kinematics and directly assess the density profile and stellar mass of the remnant. Our novel method, using distribution function models in concert with density modelling, allows us to study the GS/E remnant unhindered by contaminated kinematic selections, and represents an important step towards future 6D modelling of the entirety of the stellar halo phase space.

This paper is laid out as follows: in Section~\ref{sec:data}, we describe our APOGEE and \textit{Gaia} observations and relevant sample selection procedures used to isolate high-purity samples of GS/E stars. In Section~\ref{sec:method}, we describe the density modelling framework, paying specific attention to the integration of distribution function models. We also thoroughly validate our new approach using mock data. In Section~\ref{sec:results}, we present the results of our analysis of the GS/E remnant, including fits to the whole stellar halo for comparison. In Section~\ref{sec:discussion}, we discuss results in the context of recent literature, and also address the limitations of our new modelling technique. We end in Section~\ref{sec:summary-conclusions} by summarizing our results and looking ahead to future applications of distribution functions to studies of the stellar halo.

\section{Data}
\label{sec:data}

\subsection{Observations and stellar properties}
\label{subsec:observations}

We use data from APOGEE \citep{apogee}, specifically the sixteenth data release \citep[DR16][]{sdssdr16}, a component of the Sloan Digital Sky Survey IV \citep[SDSS][]{sdss4}. We do not use the most recent APOGEE data as we prefer to use the same data as \citetalias{lane22}. APOGEE-1 \citep[an SDSS III program;][]{sdss3} and its successor APOGEE-2 (an SDSS IV program) obtain high resolution (R $\approx 22,500$), near infrared ($1.51 \mu\mathrm{m} < \lambda < 1.7 \mu\mathrm{m}$), high signal-to-noise ($> 100$ per pixel) spectra using two near-identical spectrographs \citep{apogee_spectrographs} at the 2.5~m Sloan Telescope \citep{gunn06} at the Apache Point Observatory in New Mexico (APOGEE and APOGEE-2N) and the 2.5~m Ir\'en\'e du Pont Telescope \citep{bowen73a} at the Las Campanas Observatory in Chile (APOGEE-2S). APOGEE DR16 contains observations of approximately 430,000 unique stars from both hemispheres, which span a range in Galactocentric radii and heights above and below the disk, including all major stellar populations of the Milky Way.

APOGEE targets are selected using 2MASS \citep{2mass} photometry by a two-fold strategy, including stars targeted specifically for focused science goals and ancillary programs as well as the systematic observations of the `main' survey. Stars in the main survey are chosen based on their reddening-corrected \citep{majewski11} $(J-K_\mathrm{S})_{0}$ colour and $H$-band magnitude from 2MASS. The targeting procedure is described by \citet{apogee_targeting} for APOGEE-1, and built upon by \citet{apogee2_targeting}, \citet{apogee2n_targeting}, and \citet{apogee2s_targeting} for APOGEE-2. Data are reduced using a pipeline originally described by \citet{apogee_pipeline} but which has been built upon for recent data releases \citep{holtzman18,jonsson20}. Throughout this paper we use stellar elemental abundances, atmospheric parameters, and radial velocities from the APOGEE Stellar Parameters and Chemical Abundances Pipeline \citep[ASPCAP,][]{aspcap}. The pipeline uses a custom line list \citep{apogee_linelist} which has been updated for DR16 \citep{apogeedr16_linelist}, and a custom stellar spectral library \citep{apogee_speclib} which has also been updated for DR16 \citep{jonsson20}.

To complement the APOGEE data we use astrometry from the third data release \citep[DR3;][]{gaiadr3} of the \textit{Gaia} space telescope \citep{gaia}. We match our APOGEE data to the \textit{Gaia} DR3 catalogue using the CDS X-match service\footnote{\url{http://cdsxmatch.u-strasbg.fr/xmatch}} implemented in the \texttt{gaia\_tools}\footnote{\url{https://github.com/jobovy/gaia_tools}} python package. We specifically rely on \textit{Gaia} proper motions, as APOGEE provides more accurate radial velocities. We further use spectro-photometric distance estimates determined using the \texttt{astroNN} artificial neural network framework\footnote{\url{https://github.com/henrysky/astroNN}} \citep{leung19a,leung19b}. We specifically take those distances from the SDSS DR17 \texttt{astroNN} Value Added Catalogue\footnote{\url{https://www.sdss4.org/dr16/data_access/value-added-catalogs/?vac_id=the-astronn-catalog-of-abundances,-distances,-and-ages-for-apogee-dr17-stars}}, which are obtained from a model trained on newer APOGEE DR17 and \textit{Gaia} eDR3 data \citep[see ][]{sdssdr17}. The Bayesian artificial neural network predicts stellar luminosity using stellar spectra, and is trained on APOGEE spectra and \textit{Gaia} parallaxes. It simultaneously predicts distances and accounts for the \textit{Gaia} parallax zero-point. Distance uncertainties are typically less than 10~per~cent, much better than \textit{Gaia} parallaxes for stars further than a few kpc from the Sun.

\begin{figure}
    \centering
    \includegraphics[width=\linewidth]{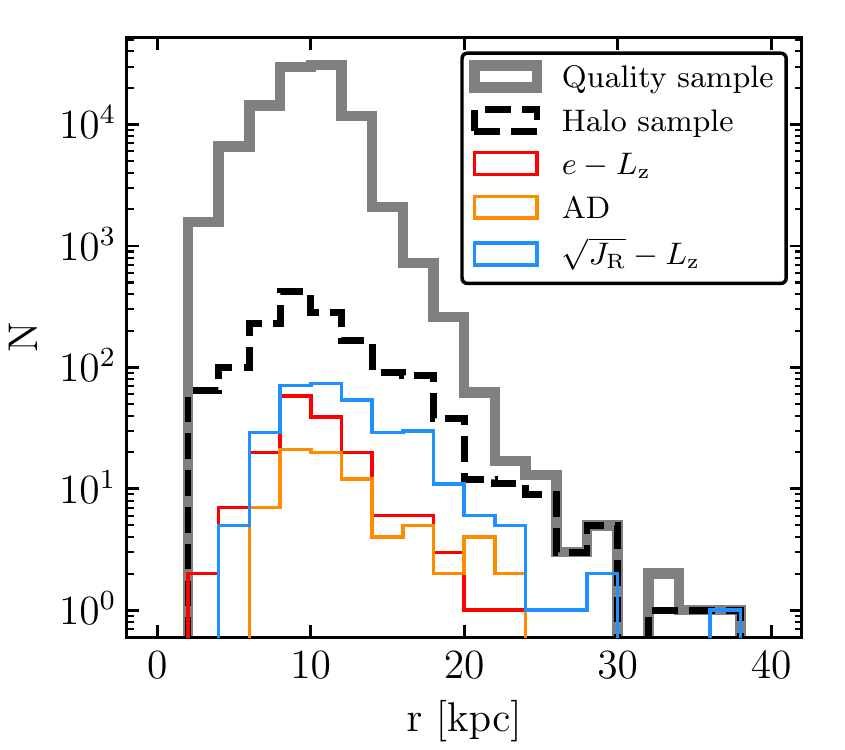}
    \caption{Distribution in Galactocentric radii of various samples used in this work. The grey line shows the high quality stars from the statistical sample. The dashed black line shows our chemically-selected halo sample. The coloured lines show the high-purity GS/E subsets of the halo sample selected using the kinematic cuts of \citetalias{lane22}.}
    \label{fig:radius_histogram}
\end{figure}

We extract a high-quality subsample of APOGEE red giants from the broader dataset. First off, we only consider stars that are part of the APOGEE statistical sample, which are stars chosen randomly according to their $(J-K_\mathrm{S})_{0}$ colour and $H$-band magnitude as described above (i.e., they are not part of some ancilliary science program, or specifically targeted for some other reason). Stars from this group are chosen which have relative distance uncertainty, $\sigma_{d}/d$, less than 20~per~cent, surface gravity between $1 < \log g < 3$, and surface gravity uncertainty less than 0.1 dex. These giants must also have measured \textit{Gaia} proper motions and positions, they must have AstroNN distances, and they must have measured APOGEE radial velocities, [Fe/H], and [Al/Fe] abundances (Note that while we will show other abundances, only these are crucial to our analysis). We also remove stars lying in fields near the Galactic bulge, and those containing stars from globular cluster so that our sample contains only stars from the smooth component of the stellar halo. We enforce the first of these criteria by removing fields within $-20^{\circ} < \ell < 20^{\circ}$ and $|b| < 20^{\circ}$. The second of these criteria is met by first identifying any fields near the on-sky location of a known globular cluster from the \citet[][December 2010 version]{harris96} catalog. For each of these fields, we examine the [Fe/H] and radial velocity distributions of the stellar constituents by eye to see if there is a noticeable enhancement of stars with properties similar to those of the cluster, paying specific attention to the region within two tidal radii of the cluster centre. Fields with a distinct group of stars with properties matching the cluster are entirely removed from consideration. Throughout the rest of the paper, we refer to this subset of APOGEE stars from the statistical sample as the `quality sample'; it contains 98,245 stars. The distribution of Galactocentric radii of this quality sample is shown in Figure~\ref{fig:radius_histogram}. The distribution is concentrated near the location of the Sun, about 8~kpc, and extends from 2~kpc to nearly 40~kpc.

We calculate kinematic properties of the stars in our sample using \texttt{galpy}\footnote{\url{https://github.com/jobovy/galpy}} \citep{bovy15}. Throughout, we assume that the Sun lies at Galactocentric coordinates $(R,z,\phi)$ = $(8.275~\text{kpc}, 0.028~\text{kpc}, 0)$ \citep{gravity21,bennett19}, the peculiar solar motion is $(U,V,W) = (11.1, 12.24, 7.25)$~km~s$^{-1}$ \citep{schoenrich10}, and that the circular velocity at the location of the Sun is 220~km~s$^{-1}$. We use a left-handed coordinate system such that angular momenta are positive in the direction of Galactic rotation. We use the \texttt{MWPotential2014} potential defined in \citet{bovy15}, and calculate energy, actions, eccentricity, and orbital extrema for each star. Actions are calculated using the `St\"{a}ckel fudge' method of \citet{binney12} whereby the Milky Way is locally approximated as a St\"{a}ckel potential using a focal length estimated with equation~9 of \citet{sanders12}. Eccentricities are calculated using a method similar to St\"{a}ckel fudge presented in \citet{mackereth18}. We principally consider three kinematic spaces of the many which are commonly used in the study of the stellar halo. First is eccentricity as a function of z-axis angular momentum (\eLz). Second is the square root of the radial action versus z-axis angular momentum (\JRLz). Third is the action diamond (AD), which has the difference between the radial and vertical actions as a function of the z-axis angular momentum, but both quantities are normalized by the absolute sum of all three actions which leads to the characteristic diamond boundary in the space. For more information on the action diamond see section~3 of \citetalias{lane22} and references therein.

\subsection{The Halo and GS/E samples}
\label{subsec:halo-gse-samples}

From the quality sample, we identify likely halo stars according to their abundances. Broadly following \citetalias{mackereth20} and \citetalias{lane22}, we focus on the iron abundance, and specifically those stars with [Fe/H] $< -1$. The principle issue with this simple selection, however, is that GS/E is known to have [Fe/H] up to $\sim -0.6$ \citep[e.g.][]{myeong19,monty20,hasselquist21,horta23a}. We therefore modify our selection to be more inclusive of GS/E, increasing our maximum [Fe/H] to $-0.6$. This introduces a different issue though, which is that a sample selected with this [Fe/H] limit would have substantial contamination from thin and thick disk populations. To mitigate this effect, we employ aluminum abundances which have been shown to reliably separate accreted halo and Galactic disk populations \citep{hawkins15}. We take specific inspiration from \citet{belokurov22} in constructing our [Al/Fe]-based selection (who also provides an illuminating discussion on the use of [Al/Fe] for this purpose), and note that several other recent works have employed similar selections, or demonstrated the separation of disk and accreted halo components on the basis of [Al/Fe] \citep{das20,hasselquist21,horta23a}. Our selection is piecewise-continuous in the [Al/Fe] versus [Fe/H] plane. At [Al/Fe] $> -0.1$, we define the halo as [Fe/H] $< -1$. At [Al/Fe] $= -0.1$, the approximate boundary between accreted and \textit{in-situ} populations \citep{hawkins15,das20,hasselquist21}, our selection increases in [Fe/H] to the locus ([Fe/H],[Al/Fe]) $= (-0.6,-0.3)$. Our selection then includes [Fe/H] $< -0.6$ for [Al/Fe] $< -0.3$.

\begin{figure*}
    \centering
    \includegraphics[width=\linewidth]{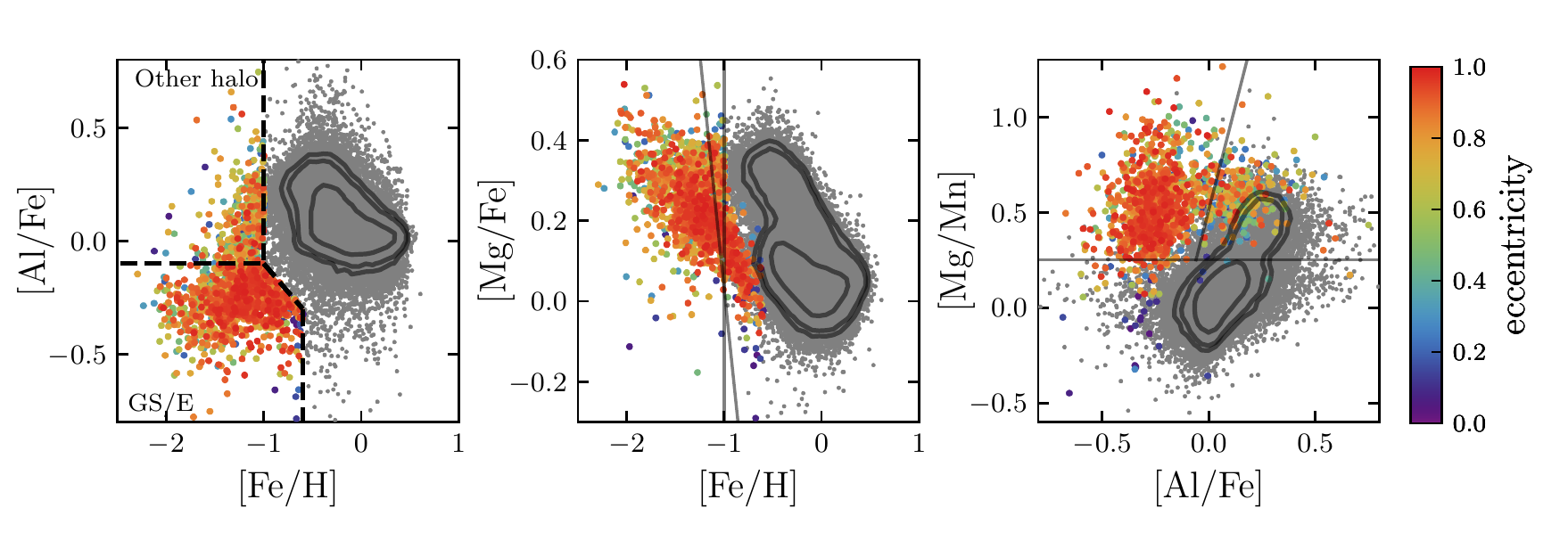}
    \caption{Abundances of the stellar sample: [Al/Fe] versus [Fe/H] (\textit{left}), [Mg/Fe] versus [Fe/H] (\textit{middle}), and [Mg/Mn] versus [Al/Fe] (\textit{right}). Coloured points show stars from the halo sample, selected chemically by their [Fe/H] and [Al/Fe] abundances. The dashed lines in the left panel show the selection used to pick these stars. Points are coloured by their eccentricity. Grey points are stars from the quality sample not chosen to be part of the halo. The thin solid lines in the middle and right panels show reference lines and selections from the literature (see text).}
    \label{fig:halo_abundances}
\end{figure*}

This selection (dashed line), and the 1,523 halo stars from the quality sample chosen with it, are shown in the left panel of Figure~\ref{fig:halo_abundances}. The radial range of this sample is also shown as a dashed line in Figure~\ref{fig:radius_histogram}; it extends about as far as the quality sample. In Figure~\ref{fig:halo_abundances}, we display the stars defined as part of the halo sample in other abundance planes commonly used in the literature. The middle panel shows [Fe/H] versus [Mg/Fe], showing a large group of stars at high [Mg/Fe] and low [Fe/H] traditionally attributed to the \textit{in-situ} halo and thick disk, accompanied by an extended group at lower [Mg/Fe] and higher [Fe/H] which is emblematic of GS/E and other accreted populations \citep[e.g.][]{hasselquist21,horta23a}. The solid lines are fiducials from other studies, the vertical line marks [Fe/H] $= -1$, which is the cutoff adopted by \citetalias{mackereth20}, and the sloped line is used by \citet{mackereth19a} and \citetalias{lane22} to separate disk and halo populations. It is clear that these cuts, which rely predominantly on [Fe/H], would leave much of GS/E excluded from the sample. The right panel shows [Al/Fe] versus [Mg/Mn], which is used by some authors to separate accreted and \textit{in-situ} populations \citep[e.g.][]{das20,horta21a,fernandez23}. The upper left region of the diagram is where unevolved populations such as those in dwarfs or in the old stellar halo live. The two other regions are where high and low-$\alpha$ evolved, \textit{in-situ} populations tend to live. Here we can see that the majority of our sample occupies the unevolved region, but some is found in the high-$\alpha$ evolved region, much of which are likely thick disk stars.

\begin{figure*}
    \centering
    \includegraphics[width=\linewidth]{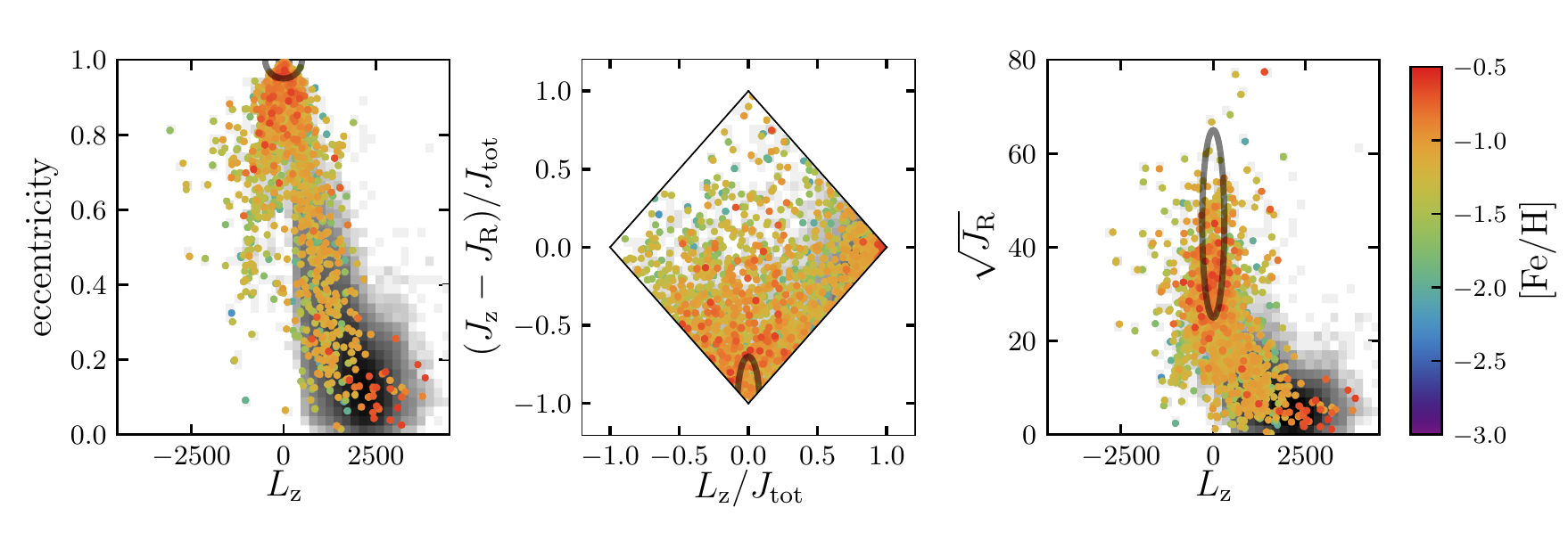}
    \caption{Kinematics of the chemically selected halo sample. Points are coloured by their [Fe/H] abundances. The thin black lines show the kinematic selections used to define the high-purity GS/E subsamples based on \citetalias{lane22}. The background histogram shows stars from the high-quality sample not chosen to be part of the halo.}
    \label{fig:halo_kinematics}
\end{figure*}

From the halo sample, we identify a subset of likely members of the GS/E remnant following the recommendations and approach of \citetalias{lane22}. Those authors built DF-based models of the nearby Milky Way stellar halo using the anisotropy parameter $\beta$ as a proxy for the two major stellar halo populations: $\beta=0.9$ for the high-[Fe/H], radially anisotropic GS/E remnant, and $\beta=0.5$ for the low-[Fe/H], more isotropic traditional stellar halo. Using these models, \citetalias{lane22} identified regions of various kinematic spaces where stars from the high-$\beta$ component were found with low contamination from the low-$\beta$ component. The `best' kinematic spaces were those where purity was around $\sim 0.8$, and these included \eLz, the action diamond, and \JRLz. We use the elliptical selection criteria from table~2 (The \textit{Survey} selections) of \citetalias{lane22}, defining a separate GS/E subsample for each of these three kinematic spaces. Figure~\ref{fig:halo_kinematics} shows these three spaces, the GS/E selections, as well as the chemically selected halo sample coloured by [Fe/H]. Our GS/E subsamples defined using the \eLz\ plane, action diamond, and \JRLz\ plane contain 163, 77, and 319 stars respectively.

We consider three kinematic spaces here so that we may examine the differences in outcomes of our model-fitting procedure. This type of selection constitutes a significant bias when modelling GS/E, in so far as we discount a large fraction of stars belonging to the high-$\beta$ GS/E population where they overlap with the low-$\beta$ halo. To wit, completeness for these selections is typically between $0.3-0.7$ and, as we shall show, is spatially non-trivial. We return to these DF-based models in \S~\ref{subsec:kinematic-effective-selection-function} when we discuss the creation of the kinematic effective selection function which addresses these biases in the density modelling framework.

The final step in cultivating our subsamples of GS/E stars for modelling is to address contamination from thick disk populations. \citetalias{lane22} include a thin and thick disk component in their models, but acknowledge that this simplifies the complicated underlying nature of the Milky Way disk, which is closer to a continuum of disk populations parametrized by scale height, length, and velocity dispersion \citep{bovy12}. The selections of \citetalias{lane22} find near-perfect separation between thick disk and high-$\beta$ populations, but their simple model is unable to account for the kinematically hottest thick disk populations. Even though these populations are numerically insignificant in the context of the density of the stellar disk, they are significant in the context of the density of the stellar halo.

\begin{figure*}
    \centering
    \includegraphics[width=\linewidth]{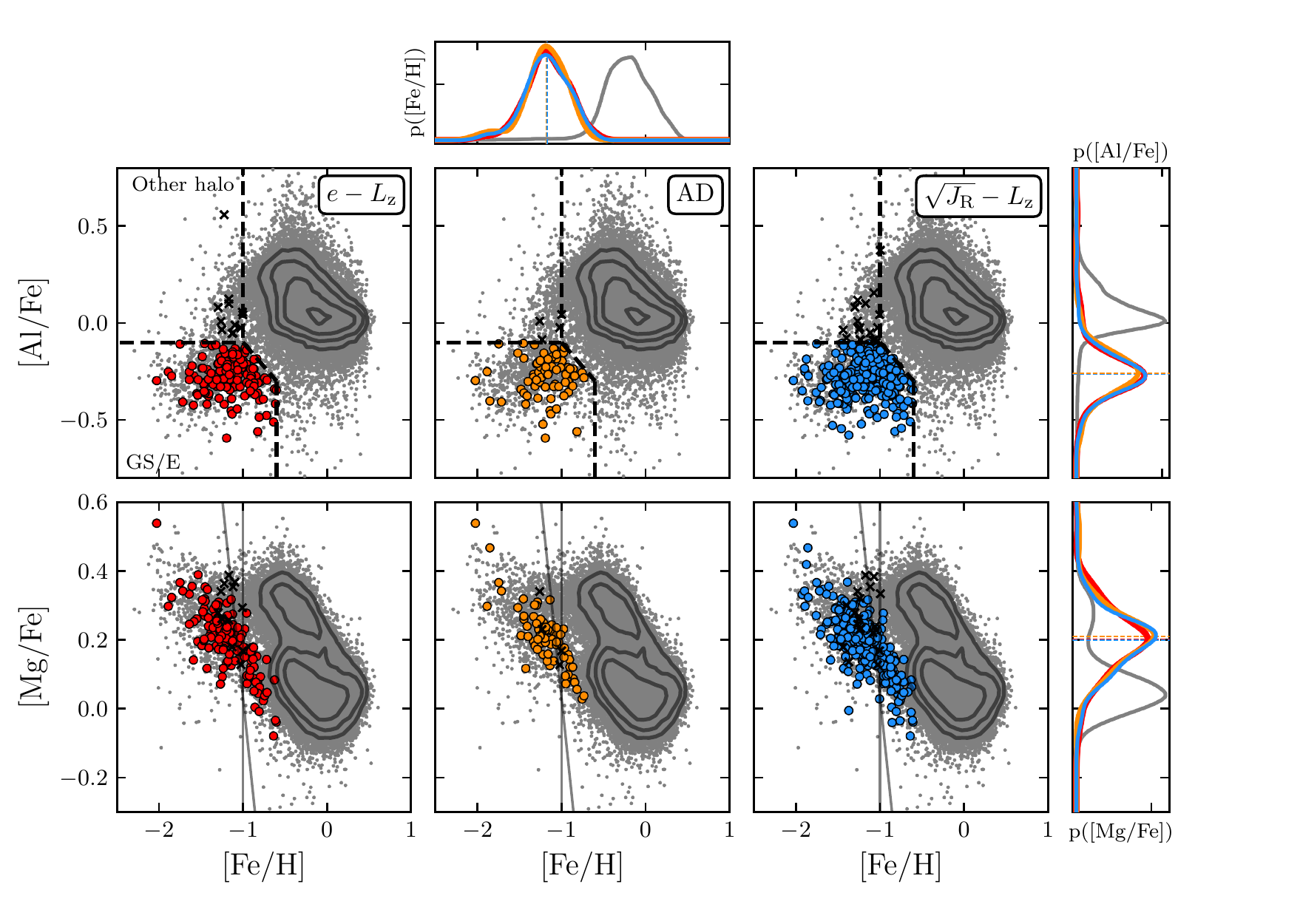}
    \caption{Abundances of the kinematically selected GS/E subsamples: [Al/Fe] versus [Fe/H] (\textit{top}) and [Mg/Fe] versus [Fe/H] (\textit{bottom}). Each column shows a high-purity GS/E subset of the stars from the halo sample chosen using one of the kinematic selections (labelled). The solid and dashed are the same as described in Figure~\ref{fig:halo_abundances}. The points which have [Al/Fe] $< -0.1$ are deemed to be accreted stars, and therefore part of the GS/E subsamples, and are shown with coloured points in both top and bottom panels. The points with [Al/Fe] $> -0.1$ are deemed to be \textit{in-situ} stars, not part of the GS/E subsamples, and are shown with black crosses. The top-most panel shows the marginalized [Fe/H] abundance for all stars (grey), as well as each of the three kinematic samples (coloured according to the main panels in the figure). The lines in this margin are Gaussian kernel density estimates. The dashed lines show the medians of the samples. The two right-most panels show the margins for [Al/Fe] and [Mg/Fe] constructed in the same way.}
    \label{fig:selection_abundances}
\end{figure*}

Fortunately, these populations should be somewhat distinct in abundance space. Figure~\ref{fig:selection_abundances} shows the GS/E subsamples in [Fe/H] versus [Al/Fe] (top row) and [Mg/Fe] (bottom row). Since \textit{in-situ} populations (both disk and stellar halo) should occupy the higher [Al/Fe] portion of our halo selection region, we make a cut at [Al/Fe] $= -0.1$ and only consider kinematically selected stars with lower [Al/Fe] to be part of our GS/E subsamples. In Figure~\ref{fig:selection_abundances}, the horizontal dashed line is set at [Al/Fe] $= -0.1$, and stars with greater [Al/Fe] abundances are shown as black crosses in both sets of panels. In [Fe/H] versus [Mg/Fe], many of the stars which are discounted occupy the high-[Mg/Fe], intermediate-[Fe/H] part of the distribution, exactly where thick disk and \textit{in-situ} halo would expect to be found. We are therefore confident that this cut serves only to enhance the purity of our GS/E subsamples and does not confer any bias. The fraction of stars removed is 5--8~per~cent of the sample depending on the kinematic space. The final GS/E subsamples defined by the kinematic selections, including this cut in [Al/Fe] contain 151 (\eLz), 73 (AD), and 300 (\JRLz) stars.

The top and two rightmost panels in Figure~\ref{fig:selection_abundances} show the marginalized [Fe/H], [Al/Fe] and [Mg/Fe] abundance distributions respectively. The coloured lines are Gaussian kernel density estimates for each of the three kinematic selections and the grey line shows the same for the whole sample. Dashed lines show the medians in the same colours, which are all similar, deviating by no more than 0.02 dex for each abundance. These medians are approximately -1.17, -0.23, and 0.20 for [Fe/H], [Al/Fe], and [Mg/Fe] respectively.

\section{Method}
\label{sec:method}

In this section we describe our approach to modelling the density of both the whole stellar halo and the GS/E subsamples. We describe the manner in which we account for the biases imposed on our samples by the APOGEE selection function and by the high-purity GS/E kinematic selections. We then outline how density profiles are normalized and total masses are derived. Finally, we demonstrate that our technique is robust and yields accurate results by applying it to mock data.

\subsection{Density Modelling}

We fit density profiles to both our halo and GS/E samples, accounting for the survey selection function, dust extinction, and the colour-magnitude density of the tracer population. We generally follow the well-established approach presented by \citet{bovy12} (also discussed by \citealt{rix13}), and then built upon and used more recently by \citet{bovy16a,bovy16b,mackereth17,horta21b} and \citetalias{mackereth20}. We describe here the method in broad detail, and encourage the reader to refer specifically to \citet{bovy12,bovy16a} and \citetalias{mackereth20} for more information.

We model the observed APOGEE star counts as a Poisson point process, with rate function $\lambda(O \vert \Theta)$. Here, the data $O \equiv [\ell, b, D, \mu_{[\ell,b]}, v_\mathrm{LOS}, H, (J-K_\mathrm{S})_{0}, [\mathrm{Fe/H}]]$ are the Galactic coordinates $\ell$ and $b$, heliocentric distance $D$, proper motions in Galactic coordinates $\mu_{[\ell,b]}$, the heliocentric line-of-sight velocity $v_\mathrm{LOS}$, the apparent $H$-band magnitude $H$, the dereddened colour $(J-K_\mathrm{S})_{0}$, and the iron abundance [Fe/H]. The vector $\Theta$ of parameters describes the density profile under consideration and is to be determined by fitting to the data. The rate function can be expressed as

\begin{equation}
\label{eq:rate}
\begin{split}
	\lambda(O \vert \Theta) & = \nu_{\star}(\vec{X} \vert \Theta) \times f(\vec{v} \vert \vec{X}) \\
	& \times \vert J(\vec{X}, \vec{v}; \ell, b, D, \mu_{[\ell,b]}, v_\mathrm{LOS}) \vert \\
	& \times \rho(H, (J-K_\mathrm{S})_{0}, [\mathrm{Fe/H}] \vert \vec{X}) \\
	& \times S(\ell, b, D, \mu_{[\ell,b]}, v_\mathrm{LOS}, H, (J-K_\mathrm{S})_{0})
\end{split}
\end{equation}

\noindent where $\nu_{\star}(\vec{X} \vert \Theta)$ is the stellar number density as a function of rectangular Galactocentric coordinates $\vec{X}$ and $f(\vec{v} \vert \vec{X})$ is the velocity distribution function to which we will return in the next section when we come to the kinematic selection function. While the velocity DF in general depends on parameters $\Theta$ fit to the data, here we use a fixed velocity DF $f(\vec{v}\vert \vec{X})$ to keep the fit computationally tractable. We address the implication of this choice later in the discussion. The function $\rho(H, (J-K_\mathrm{S})_{0}, [\mathrm{Fe/H}] \vert \vec{X})$ is the apparent density of stars in colour, magnitude, and abundance space for a given position. $\vert J(\vec{X}, \vec{v}; \ell, b, D, \mu_{[\ell,b]}, v_\mathrm{LOS}) \vert$ is the Jacobian linking observed coordinates with Galactocentric coordinates, which may be split into a spatial and velocity component as $\vert J_{1}(\vec{X}; \ell, b, D) J_{2}( \vec{v}; \ell, b, D \mu_{[\ell,b]}, v_\mathrm{LOS}) \vert$. Finally, $S(\ell, b, D, \mu_{[\ell,b]}, v_\mathrm{LOS}, H, (J-K_\mathrm{S})_{0})$ is the selection function, which depends on sky location, heliocentric distance, velocity, as well as colour and magnitude. This single selection function can be separated into a spatial component, $S_{1}$, and a kinematic component, $S_{2}$, as $S(\ell, b, D, \mu_{[\ell,b]}, v_\mathrm{LOS}, H, (J-K_{\mathrm{S}0}) = S_{1}(\ell, b, H, (J-K_\mathrm{S})_{0}) \times S_{2}(\ell, b, D, \mu_{[\ell,b]}, v_\mathrm{LOS})$. The spatial component arises from the APOGEE targeting procedure, which has been briefly discussed in \S~\ref{sec:data}. This function describes the probability with which a star with a given $H$-band magnitude and $(J-K_\mathrm{S})_{0}$ colour will be included for observation in the APOGEE statistical sample, and varies based on field. We follow the approach of \citetalias{mackereth20} when constructing $S_{1}$, and details may be found in their Appendix~A. The kinematic component of the selection function arises from our selection of GS/E stars, and its construction will be described below.

As in \citet{bovy12}, the log likelihood of the model parameters may be written in terms of the rate function as

\begin{equation}
\label{eq:log-likelihood}
	\ln \mathcal{L}(\Theta) = \sum_{i} \big[ \ln \nu_{\star}(\vec{X}_{i} \vert \Theta) - \ln \int \mathrm{d}O\ \lambda(O \vert \Theta) \big].
\end{equation}

\noindent Derivation of this likelihood uses the fact that the rate function only depends on $\Theta$ through $\nu_{\star}(\vec{X} \vert \Theta)$, because as we discussed above, the velocity DF does not depend on parameters that we fit. If the velocity DF were to depend on $\Theta$ as well, the first term would simply be $\ln [\nu_{\star}(\vec{X}_{i} \vert \Theta)\,f(\vec{v}_{i} \vert \vec{X}_i,\Theta)]$. The integral in the likelihood equation is the effective volume, which expresses for a given set of model parameters the expected (non-normalized) number of stars. For APOGEE, the effective volume can be expressed as 

\begin{equation}
\label{eq:effective-volume}
\begin{split}
	\int \mathrm{d}O \lambda(O \vert \Theta) = \sum_\mathrm{fields} \Omega_\mathrm{field} \int & \mathrm{d}D\ D^{2} \nu_{\star} (\vec{X}(\mathrm{field},D) \vert \Theta) \\
	\times & \mathfrak{S}_{1}(\mathrm{field},D) \times \mathfrak{S}_{2}(\mathrm{field},D).
\end{split}
\end{equation}

\noindent Here, the effective volume is a sum over each APOGEE field of area $\Omega_\mathrm{field}$ under consideration. This is permitted for a survey such as APOGEE which is comprised of a number of small ($< 2$~degree size), non-overlapping fields, since we can assume that the density is constant across the angular extent of each field (an assumption which we test, finding it to be valid at level of about 1~per~cent). Recall in \S~\ref{subsec:observations} we removed any APOGEE field with noticeable contamination from a globular cluster, which helps to enforce this assumption. In the integral the density is as above, but the positions $\vec{X}$ are evaluated along the line-of-sight of each field. $\mathfrak{S}_{1}$ and $\mathfrak{S}_{2}$ are the spatial and kinematic \textit{effective selection functions}. Similar to $S_{1}$ and $S_{2}$, the effective selection functions include additional information required for the modelling procedure, yet which is independent of the model parameters of interest $\Theta$. 

The spatial effective selection function $\mathfrak{S}_{1}$ is determined by marginalizing $S_{1}$ over the colour, absolute magnitude, and metallicity distribution of the tracer population, including the effects of dust obscuration. It is given by

\begin{equation}
\label{eq:effective-selection-function}
\begin{split}
\mathfrak{S}_{1}(\mathrm{field},D) = \iiint & \mathrm{d} M_{H}\ \mathrm{d} (J-K_\mathrm{S})_{0}\ \mathrm{d} [\mathrm{Fe/H}] \\
& \times S_{1}(\mathrm{field}, M_{H}+\mu(D), (J-K_\mathrm{S})_{0}]) \\
& \times \rho(M_{H},(J-K_\mathrm{S})_{0}, [\mathrm{Fe/H}]) \\
& \times \frac{\Omega_{j}(H_{[\mathrm{min},\mathrm{max}],j},M_{H},A_{H}(\ell,b,D),D)}{\Omega_\mathrm{field}}.
\end{split}
\end{equation}

Here $\rho(M_{H},(J-K_\mathrm{S})_{0},[\mathrm{Fe/H}])$ and $S_{1}(\mathrm{field}, H, (J-K_\mathrm{S})_{0})$ are as above. We have introduced the absolute magnitude $M_{H} = H - \mu(D)$, which relates to the apparent magnitude by the distance modulus $\mu(D)$, and is the preferred quantity for this integration. In the main APOGEE survey each field is divided up into $j$ bins of apparent ${H}$-band magnitude and $(J-K_\mathrm{S})_{0}$, which are each bounded by a minimum and maximum magnitude $H_{[\mathrm{min},\mathrm{max}]}$. The factor $\Omega_{j}/\Omega_\mathrm{field}$ is the fraction of observable area of the plate given the local extinction and the limiting $H$-band magnitudes for each bin. $\Omega_\mathrm{field}$ is the total area of each plate, and $\Omega_{j}$ is given by

\begin{equation}
\begin{split}
& \Omega_{j}(H_{[\mathrm{min},\mathrm{max}],j},M_{H},A_{H}(\ell,b,D),D) = \\
& \Omega(H_{\mathrm{min},j} < M_{H}+\mu(D)+A_{H}(\ell,b,D) < H_{\mathrm{max},j} )
\end{split}
\end{equation}

\noindent where $A_{H}(\ell,b,D)$ is the $H$-band extinction. Intuitively, as extinction increases stars may be extinguished into or out of a given colour-magnitude bin, and the factor $\Omega_{j}$ expresses this effect. Note that since $(J-K_\mathrm{S})_{0}$ is the dereddened colour we assume it is not impacted by extinction and so does not appear in the definition of $\Omega_{j}$. The $H$-band extinction is obtained from the dust map of \citet{bovy16a}, which is a combination of the dust maps from \citet{marshall06}, \citet{green15}, and \citet{drimmel03}. We query this dust map using the \texttt{mwdust} python package\footnote{\url{https://github.com/jobovy/mwdust}}.

In practice, this spatial effective selection function integral is approximated by Monte Carlo sampling the distribution of colours, absolute magnitudes, and iron abundances of the red giant tracer population. These distributions are sourced from a grid of PARSEC v1.2S isochrones \citep{bressan12}. The grid is spaced linearly in metal fraction with $\Delta \mathrm{Z} = 0.0001$ and a minimum [Fe/H] of -3. The ages of the grid span $10-14$~Gyr, with the minimum age being appropriate for GS/E \citep{montalban21}. We draw samples from regions of the isochrone grid bounded by $1 < \log g < 3$ weighted by a \citet{chabrier03} initial mass function. The contributions from each isochrone are weighted such that the samples are spaced linearly in [Fe/H] and linearly in age. When fitting the whole-halo sample we use a [Fe/H] range of [-3,-0.6], and when fitting the GS/E subsamples we use a slightly more metal-rich range: [-2,-0.6] to reflect approximate spread in [Fe/H] observed for GS/E \citep[see our Figure~\ref{fig:selection_abundances} or refer to e.g.][]{myeong19,hasselquist21,horta23a}. Otherwise, the fits to the halo and GS/E samples both use the same isochrone grid. The effective selection function is calculated on a grid for each APOGEE field with 300 bins between $7 < \mu(D) < 19$ (roughly $0.25~\mathrm{kpc} < D < 65~\mathrm{kpc}$) spaced linearly in distance modulus.

\subsection{The Kinematic Effective Selection Function}
\label{subsec:kinematic-effective-selection-function}

The kinematic effective selection function can be expressed as:

\begin{equation}
\label{eq:kinematic-effective-selection-function}
\begin{split}
    \mathfrak{S}_{2}(\mathrm{field},D) = \int & \mathrm{d} \vec{v}\ f(\vec{v} \vert \vec{X}(\ell, b, D)) \\
    & \times \lvert J_{2}(\vec{v}; \ell, b, D, \mu_{[\ell,b]}, v_{\mathrm{LOS}}) \rvert \\
    & \times S_{2}(\ell, b, D, \mu_{[\ell,b]}, v_{\mathrm{LOS}})
\end{split}
\end{equation}

where $f$, $J_{2}$, and $S_{2}$ are all as above, but defined at field locations $(\ell,b,D)$. As discussed by \citetalias{mackereth20}, the ideal approach for handling stellar kinematics is to have a parametrized DF $f(\vec{v} \vert \vec{X}, \Theta)$ which may be used to kinematically select populations with a set of rules $S_{2}$. The downside of this approach is twofold: first, DFs -- particularly realistic ones -- tend to be computationally expensive to compute, and second, this expands the coordinate space to 6D. These two effects combine such that it is impractical to use DFs with varying parameters which must be integrated on the fly during the optimization of the likelihood.

\citetalias{mackereth20} bypassed this issue by asserting that the two major halo stellar populations: the radially anisotropic, high-[Fe/H] GS/E remnant, and the more isotropic, low-[Fe/H] traditional stellar halo are well-separated by eccentricity. In this work, we adopt a more sophisticated approach, leveraging the DF-based models of \citetalias{lane22}, which we have already discussed in the context of GS/E sample selection. We will use this model to construct the kinematic effective selection function in such a way that it is independent of any model parameters $\Theta$.

To briefly summarize the model of \citetalias{lane22}, they consider anisotropic distribution functions of the form \citep[see, e.g.][]{binney08}

\begin{equation}
    \label{eq:anisotropic-df}
    f(\mathcal{E},L) = L^{-2\beta}f_{1}(\mathcal{E}).
\end{equation}

Here, $\mathcal{E} = \Psi - \frac{1}{2}v^{2}$ is the binding energy and $\Psi = -\Phi + \Phi_{0}$ is the negative of the gravitational potential normalized such that $\Psi(\infty) = 0$. $L$ is the total angular momentum, $v$ is the velocity magnitude, and $\beta$ is the anisotropy parameter, defined as

\begin{equation}
    \label{eq:beta}
    \beta = 1- \frac{\sigma_{T}^{2}}{2\sigma^{2}_{r}}\,,
\end{equation}

\noindent where $\sigma_{T}$ is the quadrature sum of the azimuthal and polar velocity dispersions, and $\sigma_{r}$ is the velocity dispersion in the radial direction. The function $f_{1}$ relates the mass density of the DF to the underlying potential and is, in general, non-trivial to compute, often requiring solving multiple integrals numerically \citepalias[For more information see ][ particularly Appendix A]{lane22}.

We use the same underlying potential \citep[\texttt{MWPotential2014 of }][]{bovy15} and mass density for the DF as \citetalias{lane22}. The mass density is a spherical truncated power law with power law index $\alpha=3.5$ and truncation radius $r_{c} = 25$~kpc, taken from the stellar halo fits of \citetalias{mackereth20}. We use $\beta=0.3$ to represent the low-[Fe/H] traditional stellar halo, and $\beta=0.8$ to represent the high-[Fe/H] GS/E halo. We modify our choice of high- and low-$\beta$ compared with \citetalias{lane22}, who use $\beta=0.9$ and $\beta=0.5$. A smaller value for the low-$\beta$ component aligns slightly better with the accepted values for the metal-poor halo from the literature \citep{belokurov18,fattahi19,lancaster19,iorio21}. With regards to the anisotropy of the high-$\beta$ component, the literature does support a value of $\beta \sim 0.9$ \citep{belokurov18,fattahi19,lancaster19,iorio21}, However we prefer to be conservative with our choice of $\beta=0.8$ since \citetalias{lane22} did note some minor numerical issues when using DFs with higher values of $\beta$. Additionally, as we shall show, by modelling GS/E with $\beta=0.8$ we are actually constructing an upper limit on the mass when compared with a mass derived using a higher value of $\beta$.

 We set these two components to have equal density in our models in order to enforce the assumption that the two components exist in roughly equal density near the position of the Sun. While it may seem odd that we are fixing the parameters of the mass density distribution of the stellar halo in our DF models before setting out to use them to measure the mass density distribution of the stellar halo, we are confident of a scheme to ensure that this does not significantly bias our results, and will discuss more in \S~\ref{sec:discussion}. It is worthwhile to note though that only the high-$\beta$ component is relevant for calculating the kinematic effective selection function, the low-$\beta$ component is only used to estimate purity, which is simply informative. Additionally, the overall normalization of this high-$\beta$ DF is also unimportant, only its shape, which is determined by the value of $\beta$ as well as the form of the underlying density profile. These two factors contribute to the reasoning why the kinematic effective selection function is resilient to changes in the assumptions of our models, which we will expand upon later.

\citetalias{lane22} studied samples drawn from these DFs at positions where APOGEE had observed stars. We adopt a similar approach when computing $\mathfrak{S}_{2}(\mathrm{field},D)$. At each location where we calculate $\mathfrak{S}_{2}$, we draw $10^{3}$ velocity samples from our DFs and calculate actions and eccentricities in a manner identical to that used for our stellar samples (\S~\ref{sec:data}). We then calculate the completeness of the high-$\beta$ component given each of the kinematic selections used in \S~\ref{subsec:halo-gse-samples} (i.e., the fraction of high-$\beta$ samples lying in the selection compared with the total number of high-$\beta$ samples), which is equivalent the value of $\mathfrak{S}_{2}$. Drawing samples from DFs and calculating kinematic quantities is computationally expensive, and so rather than compute $\mathfrak{S}_{2}$ for each point on a grid identically to the grid on which we computed $\mathfrak{S}_{1}$, we compute it only for 21 points, evenly spanning the same range of distance modulus (7 to 19), for each field. We then linearly interpolate the value of $\mathfrak{S}_{2}$ between these 21 points when combining it with the more densely sampled $\mathfrak{S}_{1}$. 

\begin{figure*}
    \centering
    \includegraphics[width=\textwidth]{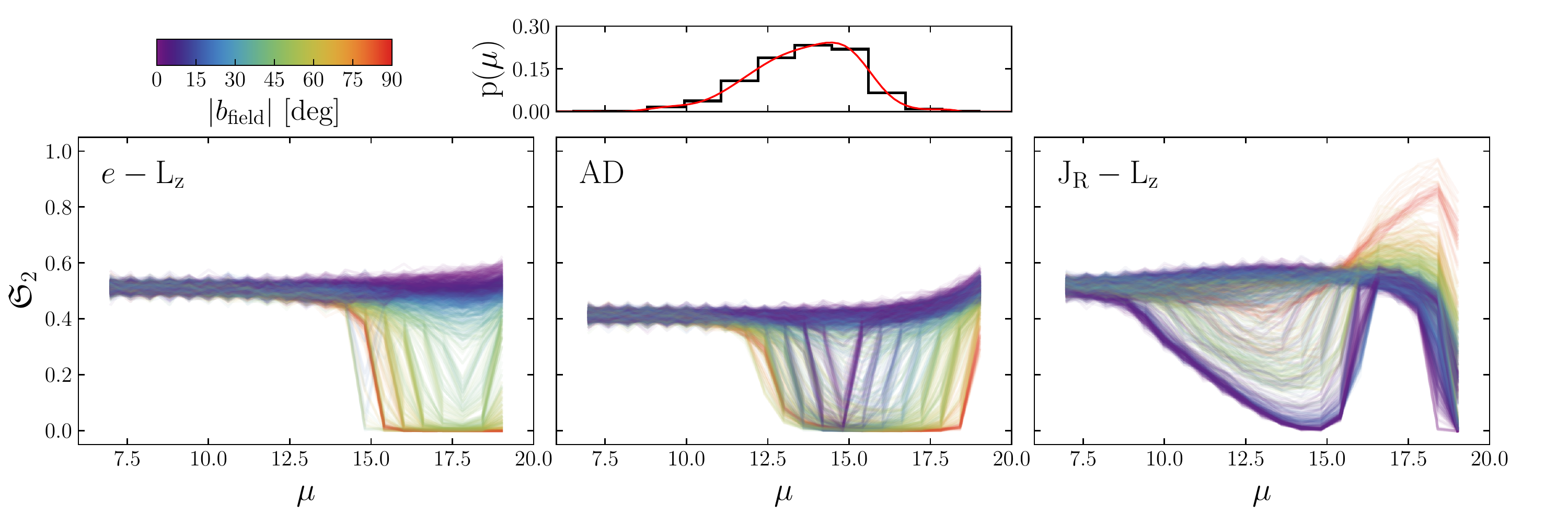}
    \caption{Kinematic effective selection function shown versus distance modulus, $\mu$, for each line of sight in APOGEE. Each of the large panels shows the kinematic effective selection function derived using a different set of kinematic parameters (labelled) and its corresponding selection criterion (see \S~\ref{subsec:halo-gse-samples}). The colour of the individual lines shows the absolute value of the Galactic latitude of the field coordinates. The small top panel shows the distribution of distance modulus for the halo subsample, and the red line shows the cubic spline fit to the distribution, which is used to marginalize $\mathfrak{S}_\mathrm{2}$ along the line of sight in Figure~\ref{fig:ksf_lb_dmod_marginalized}.}
    \label{fig:ksf_dmod_fields}
\end{figure*}

\begin{figure}
    \centering
    \includegraphics[width=\linewidth]{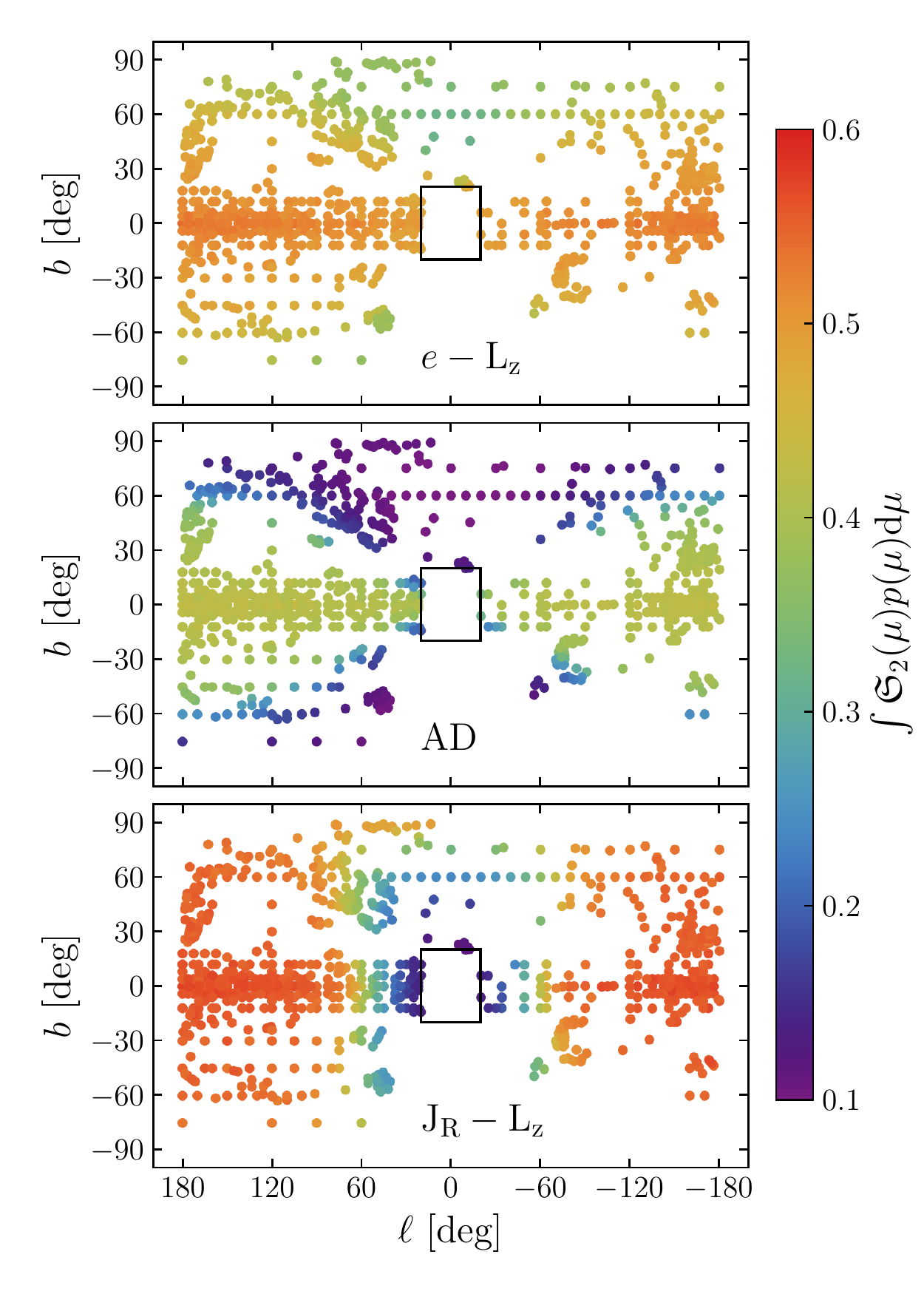}
    \caption{The kinematic effective selection function, $\mathfrak{S}_\mathrm{2}$, marginalized over the distance modulus distribution of sources (see Figure~\ref{fig:ksf_dmod_fields}) for each line of sight shown in Galactic coordinates. The three panels show three kinematic spaces used to produce the kinematic effective selection functions (labelled).}
    \label{fig:ksf_lb_dmod_marginalized}
\end{figure}

Figure~\ref{fig:ksf_dmod_fields} shows the value of $\mathfrak{S}_{2}$ as a function of distance modulus for each of the three kinematic selections. Each field is shown as a different line, coloured by the absolute value of the Galactic latitude of its on-sky location. Only fields with stars included in the halo sample are shown. In a smaller panel at the top of the figure we show the distribution of distance moduli of our halo sample for reference. The value of $\mathfrak{S}_{2}$ fluctuates along a field due to the finite number of samples drawn at each location, but this noise clearly averages out in the agregate of many fields. Figure~\ref{fig:ksf_lb_dmod_marginalized} shows the value of $\mathfrak{S}_{2}$ marginalized along each line of sight using a spline fit to the distribution of distance moduli in the top panel of Figure~\ref{fig:ksf_dmod_fields} (red line) as a function of Galactic coordinates. Again, only fields which have stars in the halo sample are shown, and the $20\degr$ box we use to exclude bulge fields is shown.

These two figures each reveal an interesting perspective of $\mathfrak{S}_{2}$. Figure~\ref{fig:ksf_dmod_fields} shows how it can vary along each line of sight, and similarly Figure~\ref{fig:ksf_lb_dmod_marginalized} shows how it can vary as a function of sky coordinates. These variations are principally due to the change in kinematics as a function of radius, which when mapped into a heliocentric frame and weighted by the distribution of halo star distance moduli yields interesting patterns. A secondary effect is that samples drawn near the Galactic poles will behave differently than those at a similar radius in the plane. Directly above the centre of the Galaxy, the St\"{a}eckel approximation is a poor one, impacting actions and eccentricities, and also the kinematic selections are ill-suited to describe the distributions which arise. While this effect is somewhat pathological, it is at least entirely self-consistent within the modelling framework, such that the data and $\mathfrak{S}_{2}$ are impacted in the same way. This gives rise to the general trend of worsening $\mathfrak{S}_{2}$ with larger absolute Galactic latitude seen in both Figure~\ref{fig:ksf_dmod_fields} and \ref{fig:ksf_lb_dmod_marginalized}. For both \eLz\ and AD, $\mathfrak{S}_{2}$ is reasonably constant at distances characteristic of our sample ($\mu \sim 10-17$), with the decreases in $\mathfrak{S}_{2}$ associated with Galactocentric radius and latitude occuring at slightly lower $\mu$ for the AD space. On the other hand, $\mathfrak{S}_{2}$ has a much more complicated behaviour on a field-by-field basis for \JRLz. 

\subsection{Stellar number density models}
\label{subsec:density-models}

We consider several models for the stellar number density, which will be fit both to the whole stellar halo sample and the GS/E subsamples. We follow the approach of \citetalias{mackereth20} in our construction of these density profiles, which are general and flexible. All of our models are triaxial and expressed as functions of an effective radius.

\begin{equation}
\label{eq:effective-radius}
m = \sqrt{ X^{\prime^{2}} + \frac{Y^{\prime^{2}}}{p^{2}} + \frac{Z^{\prime^{2}}}{q^{2}} }
\end{equation}

\noindent where $[X,Y,Z]^{\prime}$ are rectangular coordinates in the rotated frame. $p$ is the ratio of the $Y^{\prime}$ to $X^{\prime}$ axes, and $q$ is the ratio of the $Z^{\prime}$ to $X^{\prime}$ axes. Each model has a flexible orientation parametrized by three variables: $\theta$, $\eta$, and $\phi$. First, the density profile is rotated in the $X^{\prime}-Y^{\prime}$ plane clockwise by an angle $\phi$, then the $Z^{\prime}$ axis is rotated towards a unit vector $\hat{z} = \big[ \sqrt{ 1-\eta^{2} } \cos\theta, \sqrt{ 1-\eta^{2} } \sin\theta, \eta \  \big]$. So the parameter $\theta$ controls the angle of the transformed $Z^{\prime}$-axis of the ellipsoid (i.e., the axis of the ellipsoid originally aligned with the $Z^{\prime}$ axis) in the plane of the Galaxy, and $\eta$ controls the degree to which the transformed $Z^{\prime}$-axis is oriented towards the Galactic north pole. Even sampling of these three parameters evenly samples the unit sphere. 

The density profiles we consider are as follows. The first is a simple power law profile

\begin{equation}
\label{eq:single-power-law}
\nu_{\star} \propto m^{-\alpha_{1}}
\end{equation}

\noindent where $\alpha_{1}$ is the power-law index. The second is a single power-law profile with an exponential truncation scale $r_{c}$, which is expressed as 

\begin{equation}
\label{eq:exp-truncated-power-law}
\nu_{\star} \propto m^{-\alpha} \exp(-m/r_{c}) .
\end{equation}

\noindent Finally, we have broken power law models with two or three break radii which take the form

\begin{equation}
\label{eq:broken-power-law}
\nu_{\star} \propto 
\begin{cases}
m^{-\alpha_{1}} & ; m \leq r_{1} \\
r_{1}^{\alpha_{2}-\alpha_{1}} m^{-\alpha_{2}} & ; r_{1} < m \leq r_{2} \\
r_{2}^{\alpha_{3}-\alpha_{2}} m^{-\alpha_{3}} & ; r_{2} < m
\end{cases}
\end{equation}

\noindent where $r_\mathrm{i}$ are the break radii (we use $r$ to emphasize that these are radii), and $\alpha_\mathrm{i}$ are the power law indices. Together these four models span a range of complexity, and are similar to others used in the literature. Throughout the rest of the paper we refer to the single power law model as `SPL', the exponentially truncated single power law model as `SC', and the broken power law models with one or two breaks as `BPL' and `DBPL', respectively.

Each of these models is considered on its own, but also with a disk contamination model. While we are confident that we have removed most of the disk contamination using abundance cuts, we recognize that some contamination is unavoidable. The disk contamination model is exponential, with radial scale length 2.2~kpc and vertical scale length 0.8~kpc \citep{mackereth17}. We parametrize contamination by including a factor $f_\mathrm{disk}$ which expresses the fraction of density at the solar position contributed by the disk model. The combined model takes the form

\begin{equation}
\label{eq:disk-contamination}
\nu_{\star} (m) \propto (1-f_\mathrm{disk}) \nu_{\star,\mathrm{halo}}(m) + f_\mathrm{disk} \nu_{\star,\mathrm{disk}}(m) .
\end{equation}

\noindent Models including disk contamination are referred to with `+D' after the shorthand introduced above (i.e. `SPL+D').

In fitting these models, power law indices are allowed to take any value and are nominally positive (indicating decreasing density), but occasionally the exponentially truncated models have negative $\alpha$. Indices are also required to steepen in broken power law models such that $\alpha_{1} < \alpha_{2} < \alpha_{3}$. Truncation radii for both exponentially truncated and broken power laws are required to lie in the range $[2,55]$~kpc. The inner range is motivated by the fact that we exclude any fields within a $20\degr$ box around the bulge ($20\degr$ at $\sim 8$~kpc is $\sim 2$~kpc). The outer range is driven by the fact that the effective selection function essentially becomes zero for all fields at this distance. Even though our most distant star lies at nearly 40~kpc, our statistical leverage extends to the point at which the effective selection function is zero, which is about 55~kpc. The shape parameters $p$ and $q$ are defined on the interval $(0,1]$ such that the axis of the ellipsoid aligned with the $X^{\prime}$ axis (rotated frame) is always the longest. Of the orientation parameters, $\theta$ can take values between $[0,2\pi]$, $\eta$ can take values between $[0.5,1]$, and $\phi$ is defined between $[0,\pi]$. $\eta$ and $\phi$ could take wider ranges but these are degenerate with other combinations of orientation parameters as well as $p$ and $q$. Finally, the disk contamination parameter $f_\mathrm{disk}$ is defined on the interval $[0,1]$.

\subsection{Fitting density models}
\label{subsec:fitting-density-models}

To find the best-fitting set of parameters for each model, we use Markov Chain Monte Carlo to sample the posterior of the likelihood \eqref{eq:log-likelihood}. We use the affine-invariant ensemble sampler of \citet{goodman10} implemented in \texttt{emcee} \citep{foreman-mackey13}. We first optimize the likelihood function using Powell's method \citep{powell64} and then initialize one hundred walkers with a small amount of noise around that set of optimized parameters. Each walker moves $10^{4}$ steps and we remove the first $10^{3}$ steps as burn-in. We define the best-fitting parameters as the median of this set of posterior samples, with the uncertainty being the 16th and 84th percentiles. Our posteriors are complicated, however, and while we explored other more sophisticated methods of determining the best-fitting parameters and uncertainties we defer back to this approach for its simplicity. For this reason, we make our chains available online \footnote{\url{https://doi.org/10.5281/zenodo.8339417}}.

\subsection{The mass from normalization of the amplitude of the density function}
\label{subsec:mass-calculation-from-normalization}

To determine the mass of each density profile and its array of parameter samples drawn from the posterior, we follow the approach of \citetalias{mackereth20} which we describe here briefly. First, we normalize the density profiles from \S~\ref{subsec:density-models} and then integrate the density profiles over a range of Galactocentric radii. With regards to normalization, each density profile is fit such that the density is 1 at the solar position: $\nu_{\star}(\vec{X}_{\odot}) = 1$. The number of stars expected for a given density profile and parameter vector $\Theta$ is therefore the rate function of equation~\eqref{eq:rate}, which is the product of the density and the effective selection functions, integrated over the volume of the survey. This quantity is identical to the effective volume expressed in equation~\eqref{eq:effective-volume}. Dividing the total number of observed APOGEE stars to which the density profile is fit, $N_{\mathrm{obs}}$, by this expected number of stars given $\nu_{\star}(\vec{X}_{\odot}) = 1$ gives the proper normalization of the density profile in units of observed APOGEE red giants per volume, and is expressed as

\begin{equation}
\label{eq:density-normalization}
A(\Theta) = \frac{ N_\mathrm{obs} }{ \int \mathrm{d}O \lambda(O \vert \Theta) }.
\end{equation}

We then convert this number density to a stellar mass-density using the same grid of PARSEC v1.2s isochrones \citep{bressan12} used to calculate $\mathfrak{S}_{1}$, again weighted by a \citet{chabrier03} initial mass function. We calculate the average mass of giants $\langle M_{\mathrm{giants}} \rangle$ observed in APOGEE by applying the equivalent of the observational cuts (from the APOGEE targeting procedure) in $(J-K_\mathrm{S})_0$ and our imposed cut between $1 < \log(g) < 3$ to a subset of isochrone grid defined by the minimum and maximum [Fe/H] for the sample of interest ([-3,-0.6] for the halo sample and [-2,-0.6] for the GS/E samples) and finding the initial mass function-weighted mean mass of the remaining points. Similarly, we find the fraction of stellar mass in giants, $\omega$, by taking the ratio between the initial mass function-weighted sum of isochrone points within these boundaries and those outside, again for isochrones defined by the [Fe/H] range of the sample at hand. The conversion factor between giant number counts and total stellar mass is then simply calculated as 

\begin{equation}
\label{eq:isochrone-factors}
\chi \left( \mathrm{[Fe/H]_\mathrm{[min,max]}} \right) = \frac{\langle M_{\mathrm{giants}} \rangle \left( \mathrm{[Fe/H]}_\mathrm{[min,max]} \right) }{\omega \left( \mathrm{[Fe/H]}_{\mathrm{[min,max]}} \right) }.
\end{equation}

We compute this factor for each field, adjusting the limit in $(J-K_\mathrm{S})_0$ to reflect the minimum $(J-K_\mathrm{S})_0$ of the bluest bin adopted in that field. The edges in colour binning for each field are then accounted for by our integration under $\rho(M_{H},(J-K_\mathrm{S})_0,\mathrm{[Fe/H]})$ for the effective selection function. This factor ranges between 220-370~$\mathrm{M}_{\odot}~\mathrm{star}^{-1}$ depending on the colour and [Fe/H] limits. See Appendix B of \citetalias{mackereth20} for a detailed validation of this approach using \textit{Hubble Space Telescope} photometry. Combining these factors with the number density normalization, we attain the appropriate halo mass normalization at the Sun, $\nu_{0} = A(\Theta)~ \chi \left( \mathrm{[Fe/H]_\mathrm{min,max}} \right)$ for a given sample defined by a range in [Fe/H] and a set of parameters from the posterior $\Theta$. The now properly-normalized density profiles are then integrated between Galactocentric radius $2 < r < 70$ kpc. We choose this radial range to avoid over-extrapolating our fits, and to match the range used by \citetalias{mackereth20} for efficient comparison. 

\subsection{Validation of the method using mock data}
\label{subsec:method-validation}

Since we are introducing a new layer to the modelling approach outlined in the previous section by incorporating the kinematic effecive selection function, it is prudent to test the method. We have created a pipeline to generate mock APOGEE data, including kinematics, for this purpose. We begin by considering any spherical potential expressable in \texttt{galpy} to act as the density profile representing the spatial distribution of the data. We draw mass samples from a \citet{chabrier03} initial mass function such that the total mass of the samples is equal to the target total mass of the profile. We then assign these samples a radial position by sampling from the inverse cumulative mass function for the (spherical) density profile under consideration. Azimuthal and polar angles are assigned to the samples by using random sphere point picking. We scale the $Y$ and $Z$ positions by factors $p$ and $q$ respectively to make the spherical profile triaxial and rotate it by the same method described in \S~\ref{subsec:density-models}, using a rotation by $\phi$ followed by a tilt towards $\hat{z}$ defined by $\eta$ and $\theta$. In this way we generate a set of mass samples which have positions sourced from a density profile which can be made triaxial and rotated in the same manner as our candidate density profiles.

We assign stellar parameters and magnitudes to these samples with a single 10~Gyr old, $Z=0.001$ ([Fe/H] $\sim -1.3$) isochrone from the same PARSEC v1.2 grid described above using mass as the correspondance. We then remove all samples lying outside the APOGEE footprint, compute $H$-band extinction using the dust maps described above, and remove all stars with $H$-band magnitude and $(J-K_\mathrm{S})_{0}$ colours outside of the range considered by APOGEE on a field-by-field basis. For the remaining stars we apply the same selection procedure outlined in \citet{apogee_targeting} and \citet{apogee2_targeting}, randomly selecting stars in colour-magnitude bins with the appropriate probabilities to remain in the sample. We are then left with a sample of stars that would be seen by APOGEE as if it had `observed' the underlying stellar population. The repository containing the code to create these mock data is available online\footnote{\url{https://github.com/jamesmlane/apomock}}.

To test our model, we generate a mock from a stellar population with a total mass of $2\times10^{8}$~\Msun\ and a single power law density profile with the following parameters: $\alpha=3.5$, $p=0.8$, $q=0.5$, $\theta=\pi/4$, $\eta=1/\sqrt{2}$, and $\phi=\pi/6$. We choose a low mass such that the generated mock has a small number of stars (645 for the whole mock, and 89-185 for the kinematically selected GS/E analog samples). Then we may test whether our modelling approach can recover the input parameters when the number of observations is comparable to the number we see in our GS/E subsamples ($\sim 100$). To create analog GS/E subsamples the whole mock is split in half, and each half is assigned kinematics from a DF with $\beta=0.8$ and $\beta=0.3$ respectively. We apply the kinematic cuts for each of the kinematic spaces considered in this work and then perform the fitting procedure described in \S~\ref{subsec:fitting-density-models} to the kinematically selected subsamples. We also fit the whole mock without considering kinematics, a fit which obviously does not include the kinematic effective selection function. 

The results to these fits, including the total inferred mass, are shown in Figure~\ref{fig:mock_posterior}. We are able to recover most of the correct parameters of the underlying density profile within reasonable confidence intervals, and the inferred mass of the $\beta=0.8$ population (which should be $10^{8}$~\Msun, half the value of the total population) is also within the confidence interval. This not only demonstrates that our method works correctly, but that the biases induced by the kinematic selections we employ to isolate GS/E stars can be accounted for with the correct kinematic effective selection function. The masses for the GS/E subsamples are overestimated, with typical median values being about $10^{8.1}$~\Msun. The reason for this bias is that purity for the kinematic selections is not 100~per~cent, but typically close to 80~per~cent as per \citetalias{lane22} \citep[also see e.g.][]{limberg22}. In that work purity is determined for each kinematic selection criterion using kinematics for APOGEE DR16 stars assigned by the same distribution function models that we use in this work, and so the statistics should be applicable here. The impact of an impure kinematic selection is that stars from the mock low-$\beta$ component are included in the number counts for the kinematically selected population, inflating the mass. There is a potential for a smaller effect from this contamination on the density profile parameters, since the two populations are sourced from the same density profile, but the spatially-complicated nature of the kinematic effective selection function (Figures~\ref{fig:ksf_dmod_fields} and \ref{fig:ksf_lb_dmod_marginalized}) may imprint slight spatial irregularities on this contamination. Indeed, we do see some slight mismatches between input parameters and inferred parameters for the kinematically selected populations, but they are minor. The mass on the other hand should be purely overestimated by roughly a factor of the inverse purity, which is what we see: $10^{8.1} \approx 10^{8}/0.8$. We address the implications of this effect on the fittings to real data later in the discussion when we assess systematic uncertainties.

To ensure that the disk contamination model works as expected we independently analyse the same mock, but without kinematics (i.e. no kinematic cuts on the fitting sample and no kinematic effective selection function). To the mock we add stars sourced from an identical stellar isochrone and initial mass function but tracing the exponential disk density profile used in the contamination model outlined in \S~\ref{subsec:density-models}, which has radial scale $h_{R} = 2.2$~kpc and vertical scale $h_{z} = 0.8$~kpc. We add stars from this disk model sufficient to set $f_\mathrm{disk}=0.4$ (76 stars for this mock). The process to determine the correct number of stars to add is nearly identical to the process of normalizing the density profile described in \S~\ref{subsec:mass-calculation-from-normalization}, requiring calculation of the effective volume for both the halo and disk density profile, from which the relevant density at the position of Sun can be determined. We then fit the appropriate model to these mock data, and show the results also in Figure~\ref{fig:mock_posterior}. Again, we confirm that all model parameters, including $f_\mathrm{disk}$ are recovered within the confidence intervals. Note that here we did not consider contamination from a disk component in the context of our kinematically-selected mock subsamples, since we do not have a good kinematic representation of these stars \citepalias[see][ and discussion above]{lane22}.

\section{Results}
\label{sec:results}

\subsection{Fits to GS/E subsamples}
\label{subsec:fits-to-gse}

\input{tab-structural.tex}

We first perform the fitting procedure introduced above, including the kinematic effective selection functions, to each of our kinematically-defined GS/E samples, using each of the density profiles introduced in \S~\ref{subsec:density-models}. We consider each profile with and without disk contamination. The best-fitting parameters and derived masses are shown for each density profile and each kinematic selection in Table~\ref{tab:structural}. In order to gauge how well each model fits the data, we also determine the maximum likelihood by choosing the set of parameters with the highest likelihood from the MCMC chain and optimizing our likelihood function from that set of parameters, again using Powell's method \citep{powell64}. The optimized set of parameters is never far from the highest likelihood set of parameters from the posterior, and the routine always converges quickly. To compare models with different numbers of parameters, and select the best one among them, we use this optimized maximum likelihood, $\mathrm{max}(\mathcal{L})$, to calculate the standard Bayesian information criterion $\mathrm{BIC} = k\ln(n) - 2\mathrm{max}(\mathcal{L})$ \citep{schwarz78} where $k$ is the number of parameters, and $n$ is the number of stars in the specific kinematically selected sample. We also calculate the similar Akaike information criterion $\mathrm{AIC} = 2k - 2\mathrm{max}(\mathcal{L})$ \citep{akaike74}. Both quantities penalize fits which use more parameters, but the BIC places stronger emphasis on the number of stars used for fitting, while the AIC places more weight on the value of the likelihood function. Maximum likelihoods, BICs, and AICs are shown in Table~\ref{tab:likelihood}. The value of each parameter is normalized by the value derived for the single power law model for the corresponding kinematic selection. Different kinematic selections cannot be compared in the manner we have outlined since the samples are different.

\input{tab-likelihood.tex}

\subsubsection{The \eLz\ and AD samples}
\label{subsubsec:eLz-and-AD-samples}

The fits to the \eLz\ and AD samples are similar and so we describe them together first. Beginning with the power law models: for the SPL model the power-law indices are modest, lying in the range $2 < \alpha_{1} < 3$. BPL models have very shallow inner indices, which tend to be less than $\sim 1.2$. These BPL models have outer indices which are steeper than the single index found in the SPL case, however, with values tending to be greater than 3.5. This difference in index is large, indicating a steep drop-off in density at the break radius, which tends to lie between 15 and 30~kpc. The \eLz\ sample is best-fit by smaller break radii, only 15-20~kpc, while the AD sample is best-fit by larger break radii between 20-30~kpc. The uncertainties, particularly on the fits to the AD sample, are large enough that the two sets of values are broadly consistent with one another. The exponentially truncated SC models are interesting in that their power law indices are negative, indicating that density rises with radius. However they have incredibly short truncation radii, less than 10~kpc. So the density only increases slightly in the very inner halo, and then drops off quite quickly with increasing radius as the exponential term overtakes. Thus, the overall density profile looks similar to that obtained for the broken power-law models (Figure~\ref{fig:pdensity}, which will be introduced in \S~\ref{subsec:interpretation-of-results}, shows an example of these sorts of density profiles). Adding a second break to the profile increases the steepness even more, with DBPL models showing outer power law indices of $\alpha_{3} > 6$, and inner and outer break radii between 15 to 20~kpc, and 30 to 40~kpc respectively. As in the case for the BPL break radii, when factoring in uncertainties, the sets of break radii for the DBPL models are broadly consistent between the \eLz\ and AD samples.

Disk contamination appears substantial at a glance for both samples, with $f_\mathrm{disk}$ taking values between 0.3 and 0.6. However, it is important to keep in mind that $f_\mathrm{disk}$ expresses the fraction of density at the solar position contributed by the contamination model, not the fraction of contaminating disk stars in each sample. Given that the observable volume of the contaminant disk model is vastly smaller than the observable volume of any halo model for equivalent densities at the solar position (i.e. $f_\mathrm{disk}=0.5$), even moderate values of $f_\mathrm{disk}$ do not indicate substantial contamination. The relationship between $f_\mathrm{disk}$ and the fraction of contaminating stars is non-linear, and the number of contaminating stars only becomes substantial once $f_\mathrm{disk}$ becomes very close to 1. To this point is the fact that during the construction of mocks in \S~\ref{subsec:method-validation} we only had to add to stars from the disk mock at the level of about 10~per~cent of the total (76 mock disk stars added to 645 mock halo stars) to set $f_\mathrm{disk}$ to 0.4. We convert $f_\mathrm{disk}$ from an expression of density to the number of contaminating stars for our \eLz\ and AD models, finding that the range $0.3 < f_\mathrm{disk} < 0.6$ corresponds roughly to a range of disk star contamination fraction of 0.05 -- 0.1. This is in line with our expectations from the construction of mocks and indicates that disk contamination is not a major concern from a numerical perspective. A final note on this point is the fact that best-fitting density profile parameters and masses are broadly similar between profiles with and without disk contamination.

Shape parameters vary somewhat between \eLz\ and AD models, but are broadly consistent across non-SPL fits. In general, we find $p$ of order $0.5-0.7$ and $q$ of order $0.4-0.6$, which suggests significant triaxiality. For AD models the values of $p$ and $q$ are very similar, while for the \eLz\ models $p$ tends to be greater than $q$. For \eLz\ models, the value of $\eta$ is close to unity, indicating near alignment with the Galactocentric coordinate system. For density profiles such as these, where $\eta$ is close to 1, the value of $\theta$ is not particularly significant (refer to the definition of $\hat{z}$ in \S~\ref{subsec:density-models}). AD models, on the other hand, tend to have a large spread in $\eta$, with median values being around 0.8. The range of values is allowed because $p$ and $q$ are similar, causing degeneracy in rotation about the principal axis. Despite the flexibility in $\eta$, the value associated with the maximum likelihood set of parameters is typically close to 1. The final shape parameter, the azimuthal rotation $\phi$, is consistent between \eLz\ and AD samples between $90^{\circ}$ and $100^{\circ}$. 

Regarding mass, the \eLz\ sample fits are found to have typically lower masses than those to the AD sample, ranging between $10^{7.9}-10^{8.2}$ ($\sim 0.8-1.6 \times10^{8}$)~\Msun\ and $10^{8.1}-10^{8.5}$ ($\sim 1.3-3 \times10^{8}$)~\Msun, respectively. The wide range in masses for each set of samples is mostly driven by the difference between SPL and non-SPL density profiles. SPL density profiles will carry substantial mass at large radii due to the large integrable volume, while the other profiles are all broken or truncated in a way that substantially decreases the density at large radii. Discounting SPL profiles, the range of masses for the \eLz\ and AD sample fits are much narrower, being about $10^{7.9}-10^{8}$ ($\sim 0.8-1 \times10^{8}$)~\Msun\ and $10^{8.1}-10^{8.3}$ ($\sim 1.3-2 \times10^{8}$)~\Msun, respectively. The mass is not consistent between these two samples, however, differing by nearly a factor of two, although the uncertainties, of order a quarter of a decade, are large enough that the mass ranges can be reconciled with one another.

\subsubsection{The \JRLz\ sample}

While the best-fitting parameters are in reasonable agreement for the \eLz\ and AD samples, the results of the fits to the \JRLz\ sample tell a different story. Here, the power law indices tend to be somewhat steeper, and the fits using the exponentially truncated SC models do not return the type of fit described above where the power law index is negative and the truncation radius short. Instead the inner power law index is very reasonable ($\alpha_{1} \approx 1.5-2.5$) and the truncation radius is at about 35~kpc. BPL and DBPL fits do steepen, however, at break radii which are not disimilar from those found for \eLz\ and AD models, but the difference between the indices is not as stark. This difference in power law index and break radii are reflected in the derived masses, which are substantially larger for the \JRLz\ sample than the \eLz\ or AD samples. In contrast to \eLz\ and AD models, now $q$ is greater than $p$ with values ranging from $0.65-0.75$ and $0.5-0.65$ respectively. $\eta$, $\theta$, and $\phi$ are similar to the values derived for the AD and \eLz\ models, with $\eta$ being near unity, indicating only modest tilting of the ellipsoid out of the Galactocentric X-Y plane. The angle $\phi$ is between $100^{\circ}$ and $110^{\circ}$, somewhat larger than the \eLz\ and AD values, but still orienting the major axis of the ellipsoid towards the Galactocentric Y axis. Finally, disk contamination is relatively constant at about $f_\mathrm{disk} \sim 0.65$, which corresponds to a contamination fraction of $\sim 0.08$.

The masses of fits to the \JRLz samples are substantially larger than those for the \eLz\ and AD fits, roughly $10^{8.6}-10^{8.8}$ ($\sim 4-6 \times10^{8}$)~\Msun. This is largely due to the fact that the density profiles tend to be unbroken, and therefore contribute substantial mass at both large radii where the density otherwise drops off, and small radii where the density profile otherwise flattens. The masses for SPL fits are larger than broken fits, but the difference is not as marked as in the case for \eLz\ and AD fits, again since the density profiles tend to be relatively unbroken. These masses obviously cannot be reconciled with the findings for the \eLz\ and AD samples, even when factoring in uncertainties. We will discuss later in this section as well as the next why we favour the results for the \eLz\ and AD samples over these obtained for the \JRLz\ sample, and also provide explanation for why the fits to the \JRLz\ sample resemble unbroken power laws.

\subsubsection{Choice of the best-fitting density models}

\begin{figure*}
    \centering
    \includegraphics[width=\linewidth]{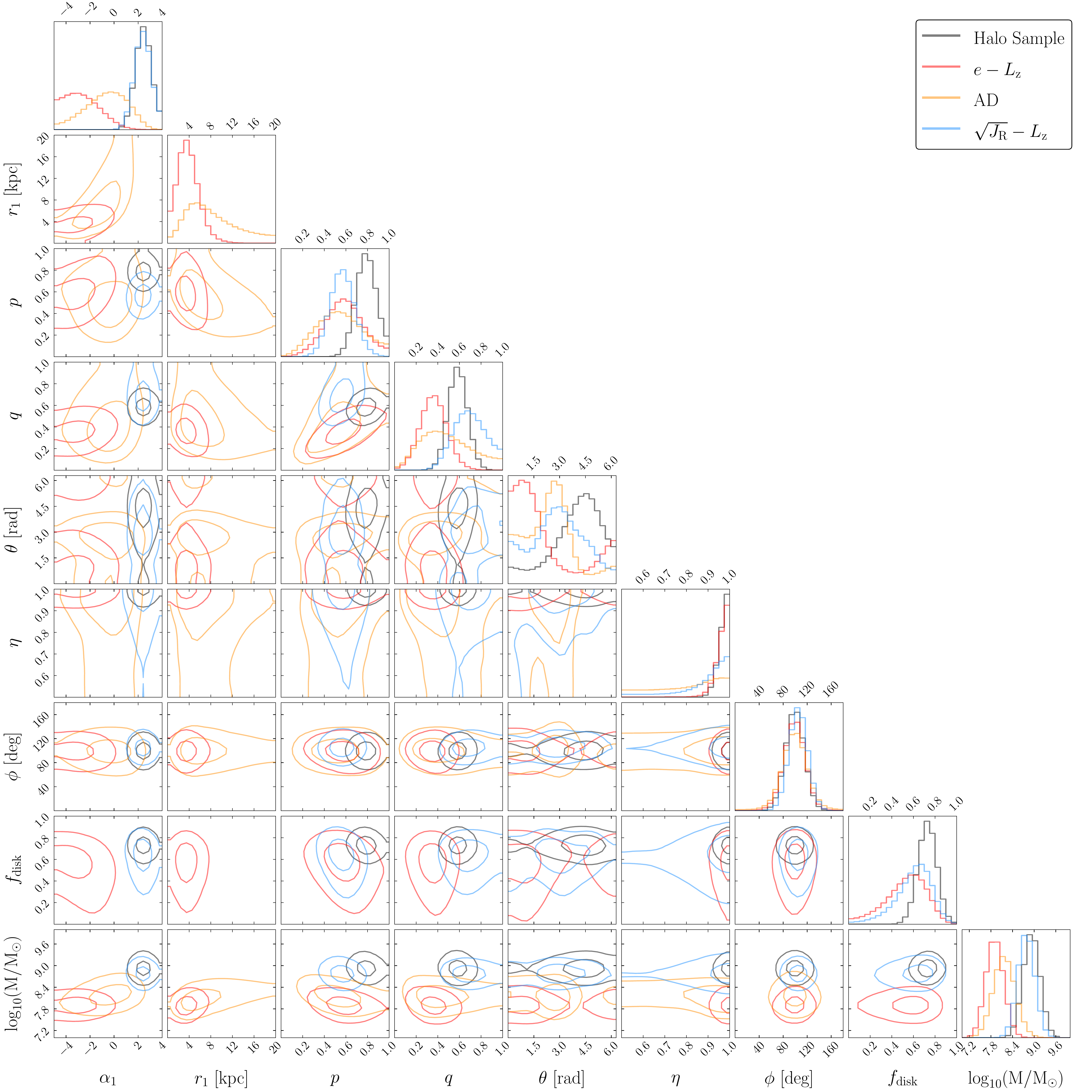}
    \caption{Corner plot summarizing the results of our best-fitting models for both the halo sample, and GS/E subsamples. Contours are missing when the best-fitting model lacks that specific parameter. The contours are at the 1 and 2 sigma levels.}
    \label{fig:posterior}
\end{figure*}

We select the best-fitting model for each kinematically-selected sample guided by the values of $\mathrm{max}(\mathcal{L})$, AIC, and BIC from Table~\ref{tab:likelihood}. Larger values of $\mathrm{max}(\mathcal{L})$ indicate a better fit, while smaller values of AIC or BIC do the same. In general, we see that maximum likelihood increases or is constant as the number of model parameters increases. The values of AIC and BIC, however, reveal that increasing the parametrization beyond SC models to BPL and DBPL models is not statistically justified. For the \eLz\ and AD samples the values of AIC and BIC peak for the SC+D and SC model respectively. For \JRLz\ AIC and BIC peak for SPL+D model. In each of these cases the maximum likelihood does not increase substantially beyond the chosen model as model complexity increases. We highlight these best-fitting models in grey in both Tables~\ref{tab:structural} and \ref{tab:likelihood}. We also show the posteriors for these best-fitting models in Figure~\ref{fig:posterior}.

We do note that even though these goodness-of-fit criteria favour simpler models over more complex models, that the more complex models tend to match the simpler models in the case of each sample. For instance, the SC and BPL models are very similar in regards to shape parameters and derived masses in the case of the \eLz\ and AD samples. While the power law and break radii appear at a glance disimilar, as mentioned above and as we shall show, these profiles are not particularly different. With this in mind, even though the simpler models are favoured from a statistical perspective, we will discuss the results of the BPL and DBPL models, particularly the break radii and different power laws.

\subsection{Fits to the whole halo}

We also performed fits using the same set of density profiles to the whole halo sample, without using any kinematic selection functions but otherwise following the exact procedure laid down in \S~\ref{sec:method}. We also calculate $\mathrm{max}(\mathcal{L})$, AIC, and BIC with the results. The best-fitting parameters and likelihood-based parameters are shown at the bottom of Tables~\ref{tab:structural} and \ref{tab:likelihood} respectively. Best-fitting structural parameters are broadly homogeneous for each of the density profiles. The inner power law index, $\alpha_{1}$, is steeper than for any of the kinematically-selected samples, tending to lie in the range 2.5-2.9. Break radii are somewhat trivial, with the posteriors hugging the upper domain limit of 55~kpc. Shape parameters are not disimilar from the \eLz\ and AD fits, although the halo is much closer to a simple oblate ellipsoid, with $p$ between 0.8 and 0.9 and $q$ taking values between 0.5 and 0.6. Orientation parameters suggest strict alignment with the Galactocentric frame, with $\eta$ very close to 1, which renders $\theta$ uninformative. $\phi$ takes values of about $100\degr$, consistent with the values found for the kinematic samples, suggesting alignment of the major axis of the density ellipsoid with the Galactocentric Y-axis. Finally, $f_\mathrm{disk}$ is higher than for the kinematic subsample fits, with best-fits occupying a narrow range 0.73 -- 0.75, which are equivalent to a numerical contamination fraction of 0.18 -- 0.2.

Masses for the whole halo are much larger than for the GS/E subsamples, lying between $10^{8.8}-10^{8.9}$ ($\sim 6.5-8.5\times10^{8}$)~\Msun\ depending on the density profile. Since the break radii lie at large Galactocentric distances, the fractional difference between SPL and non-SPL models is much less substantial than for the other kinematic subsamples. Taking the same approach to gauging goodness-of-fit as above, we can say that the best-fitting density profile is the SPL+D model.

One interesting trend is that fits to the halo sample are very similar to those for the \JRLz\ sample. Power law indices of the best-fitting models are both about 2.5, and the masses are comparable. The shape parameters are disimilar between the two best-fits, But the orientation parameters are very similar. Both $f_\mathrm{disk}$ and the mass of the \JRLz\ best-fitting profile is more similar to those of the halo best-fitting profile than either the \eLz\ or AD best-fits. All this suggests that perhaps the \JRLz\ selection is not as high in purity as suggested by \citetalias{lane22}, and the resulting sample may be closer to an intrinsic whole-halo mix of high and low $\beta$ populations, therefore yielding similar fits.

\subsection{Interpretation of results}
\label{subsec:interpretation-of-results}

\begin{figure*}
    \centering
    \includegraphics[width=\linewidth]{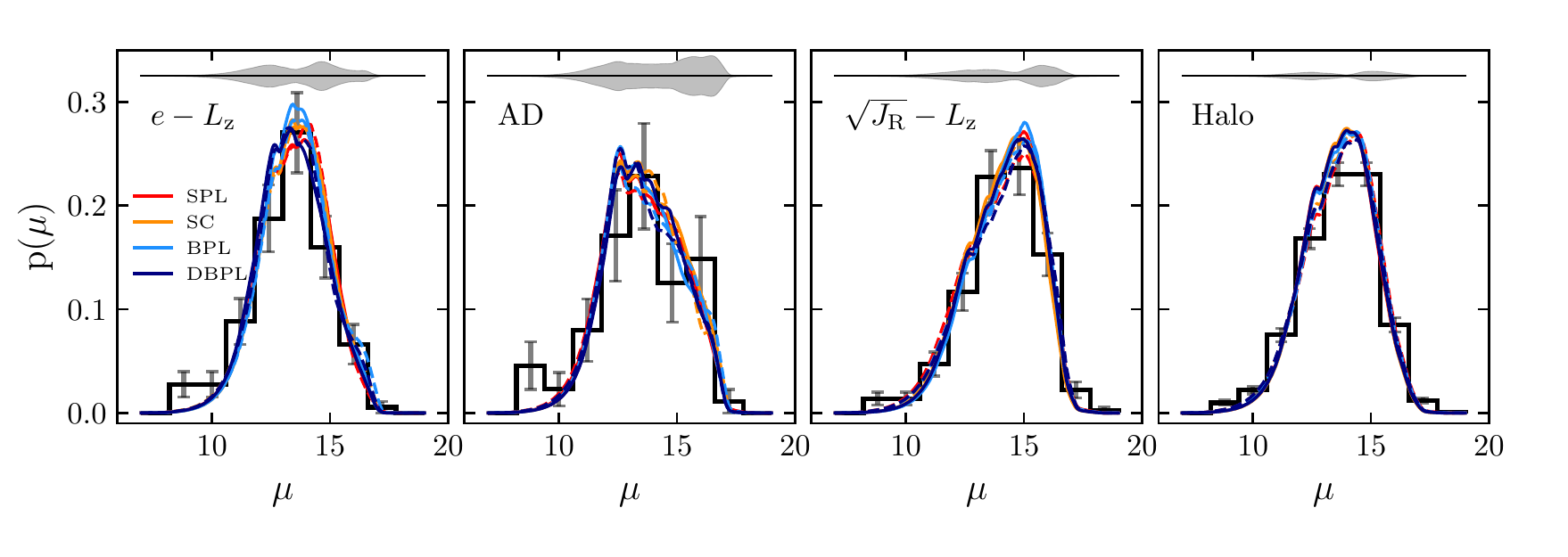}
    \caption{Histograms of distance modulus for each kinematically-selected GS/E sample, labelled by the kinematic selection used, as well as the halo sample. $\sqrt{N}$ counting uncertainty is shown as the grey error bars over each bin in each histogram. Coloured lines show median distance modulus distributions obtained by drawing random samples from the posterior distribution of best-fitting parameters. Dashed lines show density models with disk contamination, and solid lines show density models without disk contamination. The solid grey filled curve at the top of each panel shows the characteristic uncertainty in the distance modulus distribution of density profile fits (see text).}
    \label{fig:pdistmod}
\end{figure*}

Figure~\ref{fig:pdistmod} shows histograms of the distance modulus distribution for each of the samples considered here: the halo sample as well as the three kinematically-selected GS/E samples. Overlaid on each of the histograms are characteristic distributions of distance modulus from the posteriors of the best-fitting density profiles, calculated by determining the effective volume as a function of distance modulus (marginalized over all fields). These characteristic distributions are derived by taking the median of one hundred different samples from the posterior parameter distribution, selected randomly. While calculating the median of the posterior parameter samples, we also record the 16th and 84th percentile of the distribution at each distance modulus. We then take an average of these percentiles across each of the eight density profiles (SPL, SC, BPL, DBPL with and without disk contamination), which produces the characteristic uncertainty in the model distributions shown in grey at the top of each panel. Each of the density profiles are clearly able to provide reasonable fits to each of the stellar samples, with posteriors lying well within the counting uncertainty ($\sqrt{N}$ error on the counts in each distance modulus bin, shown as grey error bars) of the distance modulus distributions when also factoring in model uncertainty. There are small variations between density profiles for each sample, and particularly for density models with (dashed lines) and without (solid lines) disk contamination. 

\begin{figure}
    \centering
    \includegraphics[width=\columnwidth]{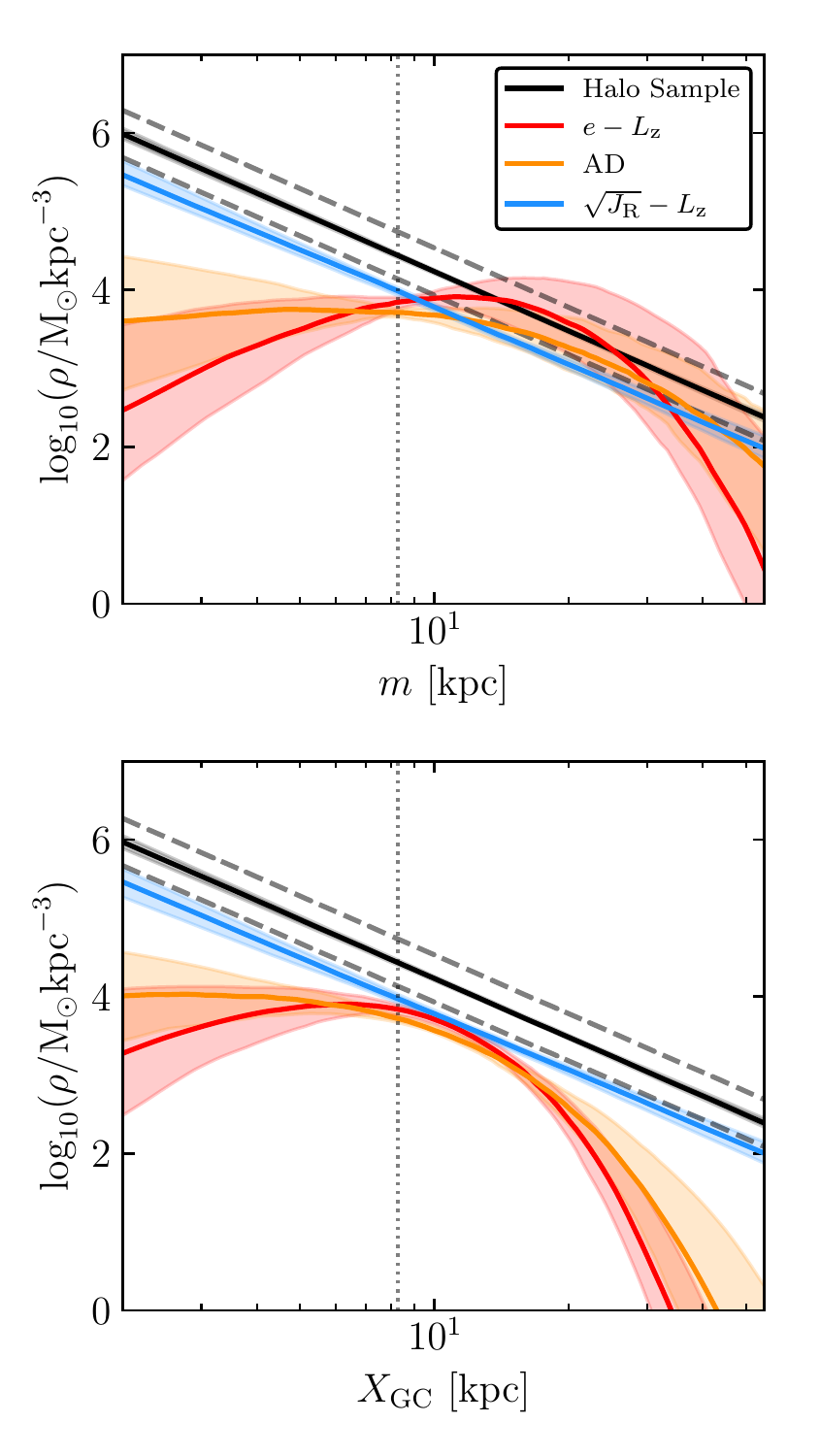}
    \caption{Mass-density posterior for the best-fitting models for the GS/E samples defined by each kinematic selection as well as the whole-halo sample. The top panel shows the density as a function of the triaxial radius $m$ \eqref{eq:effective-radius}. The bottom panel shows the density along the Galactocentric $X$ axis. In each panel the coloured bands and solid lines show the 68~per~cent interval about the median, respectively, of 100 samples of the density profile parameters drawn from the posterior. The dashed grey lines which parallel the halo profile are fiducials showing a density profile which is twice as heavy (top) and half as heavy (bottom). The dotted line in each panel shows the location of the Sun.}
    \label{fig:pdensity}
\end{figure}

Figure~\ref{fig:pdensity} shows the mass density as a function of triaxial radius $m$ (top panel) and Galactocentric $X$ (bottom panel) for the best-fitting models for each of the samples. The solid lines and the surrounding fill show the median and central 68th percentile confidence interval for the density profile generated using one hundred different sets of parameters, drawn randomly from the posterior distributions. We see emphasized here the similarities between the fits to \eLz\ and AD samples, as well as those between the \JRLz\ and whole-halo samples outlined above. The \eLz\ and AD fits are extremely flat in the inner Galaxy, even rising in the case of the \eLz\ fit, and drop-off precipitously at about $\log_{10}(m/\mathrm{kpc}) \sim 1.2-1.5 = 15-30$~kpc. The \JRLz\ sample best-fit is very similar to the whole-halo fit, a simple power law with nearly identical indices, but offset to lower density.

The dashed grey lines in Figure~\ref{fig:pdensity} shows density profiles which are half and twice as dense as the median profile for the halo sample. We show these lines because one important consistency check of our results is the observed fact that GS/E constitutes roughly 50~per~cent of stars in the solar vicinity \citep{belokurov18,lancaster19,fattahi19}. Our GS/E density profile fits do not quite rise to this level, but are not far off, sitting between one quarter to one half of the density of the best-fitting halo model within uncertainties. Interestingly, at slightly larger radii between $\log_{10}(m) \sim 1-1.5 $ or $ m \sim 10-30$~kpc the \eLz\ and AD fits do rise to -- and exceed for a small range of radii -- this level. It is useful here to compare with the density profiles as a function of Galactocentric $X$, which both factor in the effect of $p$ and $q$ and represent the density in a manner more focused on the solar neighbourhood. This panel reveals that indeed the \eLz\ and AD density profiles do peak near the solar vicinity and not beyond, but do not dominate the density at any distance.

\begin{figure*}
    \centering
    \includegraphics[width=\textwidth]{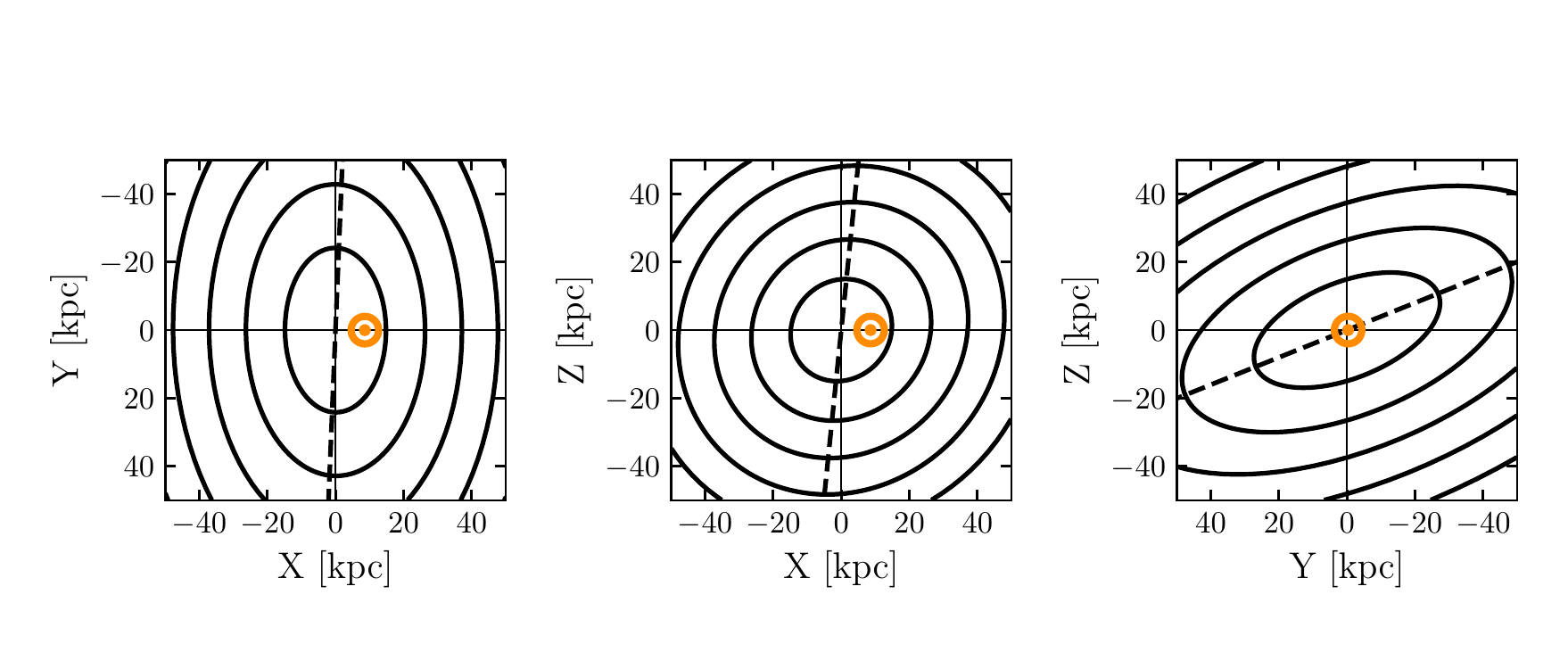}
    \caption{Contours of the best-fitting density profile (SC) for the AD sample. The parameters for the profile are the medians from Table~\ref{tab:structural}. The density is calculated in each of the X-Y, X-Z, and Y-Z planes, and so is a cross-section. The contours are set at logarithmic number densities (normalized as defined in \S~\ref{subsec:density-models} such that $\nu_{\star}(\mathrm{R}_{\odot},0,z_{\odot})=1$) of $\log_{10}(\nu_\mathrm{\star}) = [-4.5,-3.5,-2.5,-1.5,-0.5]$. The dashed line shows the principal axis of the 3D ellipsoid, projected on to each plane; it is not guaranteed to align with the cross-sectional density ellipsoids shown in the Figure. The Sun is marked with an orange circle. }
    \label{fig:pcontour}
\end{figure*}

Figure~\ref{fig:pcontour} shows density contours for cross-sectional views of the best-fitting AD profile. We show only one set of density contours, since the approximate shape of the best-fitting profiles is similar among each of the samples (i.e., triaxial, slightly tilted, and with the major axis oriented near the Galactocentric Y-axis). We can now see the density profile described by the best-fitting parameters outlined in in the previous sections. The principal axis of the ellipsoid is rotated towards the Galactocentric Y-axis ($\phi \sim 90^{\circ}$), and is tilted slightly out of the Galactic X-Y plane ($\eta$ near 1). Note that profiles where $\eta$ is equal to 1 will simply have the principal axis in the X-Y plane, but oriented still towards the Y-axis. The inclination of the principal axis with respect to the Galactocentric X-Y plane for the AD best-fitting profile shown here is $-16.4\degr$ (The central 68~percentile interval for this value is $[-35.5\degr,6.6\degr]$).

Moving to the mass, in general we find evidence for a light GS/E remnant, roughly $0.8-1.6\times 10^{8}$~\Msun\, in two out of our three kinematic samples under consideration: those created with the \eLz\ and AD selections. The \JRLz\ sample fits suggests the opposite, that GS/E is heavy, roughly $5\times 10^{8}$~\Msun. We posit that this is an issue of contamination, arising from the lower purity of the \JRLz\ selection. Indeed, \citetalias{lane22} find that the \JRLz\ is the least pure among the kinematic spaces that they studied which we also use here. To further explore this angle, we determine the purity of each of our selections in a manner similar to how we calculated the kinematic effective selection function in \S~\ref{subsec:kinematic-effective-selection-function}. The purity here is defined as the fraction of samples within each kinematic selection from the high-$\beta$ component with respect to the total number of samples in the selection, from both high- and low-$\beta$ components. We marginalize along each line of sight and over each field, weighted by the effective selection function (which generally describes the number of stars expected at a given location, modulo the density). We find the characteristic purity for the three spaces is as follows: \eLz: 0.84, AD: 0.90, and \JRLz: 0.78. This result is similar to the purity quoted by \citetalias{lane22} for each space, which was derived using similar models, albeit in a different manner.

A lower purity would impact the results to the fits in two ways. First, more stars would be counted in the fitting sample, which would drive up the normalization factor for the density profiles in equation~\eqref{eq:density-normalization}, which in turns drives up the calculated masses (see discussion of this effect in \S~\ref{subsec:method-validation} in the context of our mock data). A second, more subtle, effect is that we would expect the fit to tend more closely to the fit to the whole halo sample. The best-fitting density profile for the halo sample is steep in the inner part of the Galaxy, and lacks a cutoff at 20 to 40~kpc such as the \eLz\ and AD best-fitting density profiles have. Both of these effects contribute to a heavier overall mass, and if the \JRLz\ profile were to trend towards that sort of fit due to lower purity in the underlying sample then we would expect the mass to come out large for that reason as well. Although, it may be argued that the difference in purity (of order 6 to 12~per~cent) between the \JRLz\ and \eLz/AD spaces could not possibly account for this difference. We also acknowledge that the \JRLz\ space is `absolute' in the sense that the dependant variable is the radial action. This is contrasted to the AD and \eLz\ spaces which are much more regular, the AD space being explicitely normalized by the total action, while the \eLz\ space being normalized in the sense that eccentricity may only vary between 0 and 1, and our selection includes eccentricity of 1, which orbits approach as they become increasingly radial. The kinematic effective selection functions are derived assuming a potential, specifically \texttt{MWPotential2014} of \citet{bovy15}, but it is applied to real data. Now the actions and conserved quantities of the real data are also derived assuming the \texttt{MWPotential2014}, however their underlying kinematics are dictated by the true potential of the Galaxy, whereas the kinematics of the DF samples used to compute the kinematic effective selection function are dictated by \texttt{MWPotential2014}. This mismatch could cause problems for spaces such as \JRLz, which will be much more sensitive to the underlying potential than the AD or \eLz\ spaces.

Interestingly, the works of \citet{feuillet20} and \citep{carrillo23} suggest the opposite of the conclusion at which we have arrived here, and they posit that \JRLz\ is the superior space for selecting GS/E debris. \citet{feuillet20} conclude this by studying the metallicity distribution functions of kinematically selected halo stars, although they do not consider eccentricity or a space similar to the action diamond, finding that \JRLz\ is better than energy versus $L_\mathrm{z}$, the Toomre diagram, and radial versus tangential velocities \citepalias[this is actually consistent with the conclusions of ][]{lane22}. So while they determine that \JRLz\ is better than energies and velocities we should not take that to mean that eccentricities and the action diamond may not be better than \JRLz. \citet{carrillo23} come to their conclusions while studying simulated GS/E remnants in a cosmological context. They consider all of the spaces studied by \citetalias{lane22} but use different specific kinematic selections in some instances. For example they consider eccentricities greater than 0.7 (our eccentricity selection which is very close to 1), and a slightly larger selection in the action diamond. Their findings regarding eccentricity are broadly consistent with \citetalias{lane22}, who also find that $e > 0.7$ yields poor purity (again, note the eccentricity selection used in this work is much more restrictive). With respect to the action diamond, their results are tougher to reconcile with those of \citetalias{lane22} since the selections are similar, although their selection is again broader. In general, the work of \citet{carrillo23} benefits from knowledge of the ground truth and the realistic dynamics innate to simulations, however they lack the necessary abundance information in their simulations with which to make cuts which are typical in observations. So while perhaps their work does not discount our conclusions about the purity of kinematic spaces it certainly highlights the need to take care when kinematically selecting GS/E debris. Finally, the work of \citet{donlon23} suggests that \JRLz\ may be contaminated at high $\sqrt{J_\mathrm{R}}$ at the level of $\sim 1/3$ or greater by non-GS/E accretion remnants, which is in rough agreement with the findings of \citetalias{lane22}.

\subsection{Altering the kinematic effective selection}
\label{subsec:altering-kinematic-effective-selection-function}

When constructing the kinematic effective selection function, we noted the inconsistency between the fact that we had to assert a density profile for the stellar halo when defining the DFs, yet we are employing the DFs to determine the density profile of the stellar halo. To investigate the impact of this assumption, we compute new high and low-$\beta$ DFs based on the density profiles that we have determined for the GS/E and whole-halo samples. For the low-$\beta$ component we construct a density profile inspired by the best-fitting SPL model to the whole-halo sample. We use a single power law density profile with index $\alpha=2.5$, normalized such that it has a mass of $6\times10^{8}$~\Msun\ between $2-70$~kpc (i.e. roughly the mass of our whole-halo fit less our mass derived for the GS/E remnant). For the high-$\beta$ component we take our cue for the density profile from the BPL fits, which for the AD and \eLz\ samples have approximate inner and outer power laws of 0.8-1.2 and 3.8-5.2, and break radii between 17-27~kpc. We therefore conveniently approximate the high-$\beta$ component using a Hernquist density profile \citep{hernquist90}, which has an inner power-law index of 1 and an outer index of 4, with scale radius 20~kpc and a mass between 2-70~kpc of $1.5\times10^{8}$~\Msun. We choose to model the high-$\beta$ density profile in this way for ease of computation, and the BPL models which inspire it provide fits to the data with equivalent likelihood compared with the SC model which is our claimed preference. We keep the values of $\beta$ fixed at 0.3 and 0.8.

Samples are drawn and kinematic quantities are computed as outlined in \S~\ref{subsec:kinematic-effective-selection-function}. Now, however, we compute purity by counting the number of high-$\beta$ samples in each selection with respect to the total number of samples, but we weight the contributions from the low and high-$\beta$ components by the local density of the respective density profile (note the completeness of the high-$\beta$ component is independent of the local densities of the two components). We replicate Figures~\ref{fig:ksf_dmod_fields} and \ref{fig:ksf_lb_dmod_marginalized}, which show the kinematic effective selection functions, as Figures~\ref{fig:ksf2_dmod_fields} and \ref{fig:ksf2_lb_dmod_marginalized}, which are shown in Appendix~\ref{ap:modified-ksf}. Examining the new kinematic effective selection function (which we denote at $\mathfrak{S}_{2}^{\prime}$ in these figures) reveals remarkable consistency between them and the original kinematic effective selection functions shown in Figures~\ref{fig:ksf_dmod_fields} and \ref{fig:ksf_lb_dmod_marginalized}. On Figure~\ref{fig:ksf2_dmod_fields} we overlay the median trends, calculated across all fields at each distance modulus, of each of the kinematic effective selection functions to summarize the differences. The overall typical value of $\mathfrak{S}_{2}$ decreases slightly for each selection, from $\sim 0.5$ to $\sim 0.45$ for the \eLz\ space for example, but the overall form remains very similar. This implies that we could expect very similar fits to the data using these new kinematic effective selection functions, except that the derived mass should increase by a small amount, since the typical value of $\mathfrak{S}_{2}$ is somewhat smaller. We compute the ratio between these two median trends, and calculate an average over the range of distance modulus weighting by the distribution of distance moduli in the sample (top panel in Figure~\ref{fig:ksf2_dmod_fields} which is the same as the top panel of Figure~\ref{fig:ksf_dmod_fields}). We find that the resulting typical fractional change in the kinematic effective selection function is -0.09 (\eLz), -0.07 (AD), and -0.13 (\JRLz). We would therefore expect the masses to increase by a small amount, roughly 10~per~cent, but overall this effect is negligible. In general, these results suggest that the DF-based corrections we derive and apply in our modelling framework are resilient to modest changes, and therefore that as long as the assumptions underlying the models used to generate the corrections are realistic that their specific form is generally unimportant.

The most interesting aspect of this analysis is actually in examining purity. Above we noted that the purity is lower for the \JRLz\ selection than the \eLz\ or AD selections, but not by a significant amount (0.78 versus 0.84 and 0.9 respectively). We recalculate the typical purity using the new DFs here, again by summing the purity over all fields and distances weighted by the value of the effective selection function. We find that the new purity values are 0.78 for the AD selection, 0.65 for the \eLz\ selection, and 0.52 for the \JRLz\ selection, so all values have been lowered. Note that the overall lowering of the purities here will be largely driven by the fact that the low-$\beta$ density profile is quadruple the density at the solar position compared with the high-$\beta$ component. This should not be taken as indication that these selections are now inferior, and they are likely still optimal (i.e. were one to repeat the analysis of \citetalias{lane22}). The differences in purity among the kinematic spaces are even more stark now than in the case of the original DF models, and suggest that \JRLz\ is truly an inferior kinematic space for which to select GS/E samples. Fractionally, the \JRLz\ space experiences the largest reduction in purity, decreasing 33~per~cent. The fractional decrease for the \eLz\ and AD spaces are smaller in magnitude, 22 and 14~per~cent respectively. So not only does the AD space maintain the highest purity, but its purity is changed by the smallest amount when pivoting from the original DF models to those based on our results. These findings suggest that the AD space, in particular, should be preferred given its high purity and resilience of its purity to changes in DF models.

\subsection{Our preferred results for the density profile of GS/E}

We close out this section by summarizing our main findings for the density profile and mass of GS/E in light of our primary results and secondary analyses. We have already noted that there appears to be broad consensus between the fits to the \eLz\ and AD samples, a consensus which differs significantly from the results of the \JRLz\ sample. We also noted that the results for the \JRLz\ sample are similar in many ways to the results for the fits to the whole halo. When factoring in the assessment of purity, for both the original kinematic effective selection functions and the second set of kinematic effective selection functions derived based on our initial results, we arrive at a neat conclusion. The \eLz\ and AD results more faithfully trace the underlying reality of the GS/E remnant, while the \JRLz\ results are biased by excessive contamination from the low-$\beta$ halo such that the density profile parameters are similar to those for the whole halo and the mass is overestimated. For this reason we defer to the former results, and specifically those for the AD sample since it has the highest purity, when crafting our conclusions regarding the density profile and mass of GS/E.

With this in mind we can say that the best-fitting density profile for the GS/E remnant is the SC model, a single power law with exponential truncation, with $\alpha_{1} = -0.57^{+1.58}_{1.95}$ and $r_{1} = 8.57^{+12.58}_{-4.36}$~kpc. This type of density profile, which gently rises in the inner Galaxy and drops rapidly at modest Galactocentric radii, is non-traditional. For this reason we also report the power law indices and break radius of our BPL fit (no disk contamination) which has inner power law $\alpha_{1} = 1.05^{+0.86}_{-0.72}$, outer power law $\alpha_{2} = 3.79^{+2.44}_{-0.98}$, and break radius $r_{1} = 23.59^{+15.55}_{-9.30}$~kpc. These two profiles are, in practice, very similar over the range of radii probed and are equivalent in terms of maximum likelihood, yet the SC model is favoured by BIC/AIC since it has one fewer parameter. We find that GS/E is moderately triaxial, with the ratios of the Y-X and Z-X axes in the rotated (not Galactocentric) frame being respectively $p = 0.54^{+0.22}_{-0.2}$ and $q = 0.46^{+0.27}_{-0.2}$. The principal axis is rotated towards the direction of Galactic rotation with $\phi = 99.5^{+13.9}_{-14.9}$ degrees, and is tilted slightly out of the Galactic plane with $\eta=0.84^{+0.12}_{-0.22}$ and $\theta = 147.0^{+43.5}_{-95.3}$ degrees, which equates to an inclination of the principal axis of $-16.4^{+23.0}_{-19.1}$ degrees. Finally, the mass is $1.45^{+0.92}_{-0.51}\times10^{8}$~\Msun. 

We also calculate a few supplementary parameters that derive from our best-fitting density profile and associated mass. First, we consider the triaxiality parameter, $T$, commonly given as

\begin{equation}
    T = \frac{1-(b/a)^{2}}{1-(c/a)^{2}}
\end{equation}

\noindent where $b = \mathrm{max}([p,q])$ and $c = \mathrm{min}([p,q])$ in the context of our model parameters, and $a=1$. We calculate this parameter for each of the samples in the posterior for the SC fit to the AD sample, arriving at a value of $T=0.79^{+0.13}_{-0.35}$. This indicates that the GS/E density profile is preferrably prolate, but only to a modest degree. Second, we take our stellar mass and determine the corresponding dark matter halo mass using a stellar mass--halo mass ($M_{200}$) relation. According to the relation derived by \citet{behroozi13} (and validated by \citealt{read17} for stellar masses $\sim10^{8}$~\Msun) a stellar mass of $1.5\times10^{8}$~\Msun\ implies a halo mass of about $5\times10^{10}$~\Msun\ at redshift $z=0$, with halo masses larger by a few tenths of a decade expected for the same stellar mass at redshift $z=2$ (the approximate GS/E merger epoch). The Milky Way-oriented stellar mass--halo mass relation of \citet{nadler20} predicts a comparable halo mass of $6\times10^{10}$~\Msun. Given that the Milky Way at the time of the merger had a halo mass that is approximately half of that today \citep[so $\approx 5\times 10^{11}$~\Msun; e.g.][]{mackereth18a}, this means that the GS/E merger was approximately 1:8 to 1:10, constituting a minor merger.

\section{Discussion}
\label{sec:discussion}

\subsection{The distribution function models and the kinematic effective selection function}
\label{subsec:df-models-ksf-discussion}

Our results presented here are perched upon the DF-based models introduced by \citetalias{lane22}. While we have gone to great lengths in this paper to demonstrate that the corrections we derive for kinematic selections allow recovery of the properties of the underlying population, we must address the limits of these types of models. First, the functional form of our DFs is a simple, but flexible, \textit{ansatz} \eqref{eq:anisotropic-df} which is designed to incorporate many of the basic features we expect from stellar halo populations: simple radial dependency and a variable anisotropy which is assumed to be radially constant. It lacks, however, many of the ingredients that we know stellar halo populations to possess. For example, we know that GS/E is likely to have modest net rotation \citep{helmi18,lancaster19,iorio21}, and to have asymmetric or complicated phase-space features when comparing the debris at high and low $L_\mathrm{z}$ \citep{helmi18,naidu20}, or at large Galactocentric radii \citep{chandra23}. Additionally, as we have shown here and as has been shown in nearly every other work that has measured the density profile of GS/E, the remnant is triaxial \citep{iorio19,han22}. None of these features are included in our DF models (but of course our density models are triaxial), which may result in discrepancies between our corrected results and the true GS/E population which are challenging to predict. Finally, the mathematical form of the DF implies divergence when $\beta > 0$ as $L \rightarrow 0$, which may compromise its suitability for modelling purposes \citep[as discussed by ][]{binney14}. These effects may be particularly exacerbated when modelling the highly anisotropic GS/E remnant.

Despite these admitted deficits, our models do an excellent job of reproducing the basic assumptions and features of GS/E, namely that it is a stellar population with extreme radial anisotropy and near-zero $L_\mathrm{z}$ \citepalias[see figure 4 in][for example]{lane22}. It is likely that our models are closer to a true description of the phase-space density of GS/E than not, and therefore that our results should be taken at face value. \citetalias{lane22} tested their models against changes in underlying potential, and they also tested the resiliency of their selections to changes in radius, finding that neither change produced a noticeable effect on the interpretation of their results. Those findings may be carried over to our work to suggest that our results would be resilient to changes in underlying potential as well, while variations the quality of the selection with radial position are explicitely handled by our spatially-dependent kinematic effective selection function. Additionally, we explored variation in the kinematic effective selection function and found that at least one of our kinematic selections, AD, is resilient to iterative improvements in the selection function based on our initial results. These tests all suggest that our preferred findings of a light, triaxial GS/E are accurate and would not change substantially given more sophisticated corrections based on more realistic kinematic models.

Despite our confidence in these results, we do plan follow-up work which will both help to reinforce our findings, but also to expand these sorts of techniques for use with other remnants, which tend to have more complicated kinematic properties, requiring more complicated kinematic selections and corrections during the modelling procedure. A tangential, yet perhaps more realistic, approach would be to model the stellar halo as a mixture of self-consistent DFs \citep[see, e.g.][]{lancaster19,iorio21} for similar studies using Gaussian mixtures), however the large number of parameters and the intractability of computing such DFs on the fly mean that it would be somewhat challenging. On a similar note, exploring the use of more realistic DFs should be of top priority. For example action-based DFs \citep{binney14,posti15,williams15} including flattening and rotation \citep{binney14}. The construction of self-consistent triaxial DFs is an area of ongoing effort \citep{sanders15,binney18}, yet these would obviously prove ideal given the observed geometry of the GS/E remnant.

Another fruitful avenue for future work would to test these sorts of modelling techniques on merger remnants in simulated Milky Way analogs sourced from cosmological simulations. The EAGLE \citep{schaye15}, AURIGA \citep{grand17}, and FIRE \citep{wetzel16} sets of simulations, among others, all have rich veins of recent work on GS/E-like remnants \citep[e.g.][]{fattahi19,mackereth19a,horta23b,orkney23}. In particular, examining the phase space distributions of debris in detail would be useful. There have been recent works that have looked at the phase space distribution of simulated debris \citep[e.g.][]{amorisco17,jean-baptiste17,naidu21,amarante22,orkney23,carrillo23}, and we advocate a more in-depth look specifically at how well various families of DFs can be fit to merger debris. This type of study could place necessary constraints or priors on the vast volume of parameter space which would undoubtedly accompany a multi-DF assessment of the Milky Way stellar halo.

\subsection{The measurement of the mass of the GS/E remnant}

The core measurement made in this paper is the stellar mass of GS/E, which we estimate to be about $1.5\times10^{8}$~\Msun\ in our best-fitting models. We will discuss our results in the context of the literature in the next section, but so as to not bury the lead: our derived mass for GS/E is much lower than other findings. It therefore behooves us to assess our measurement of the mass of GS/E and investigate possible sources of systematic error.

First, we examine the impact of a few of our sample-selection and model-fitting decisions on our final results. We repeat our fits without the $\log g$ uncertainty and relative distance uncertainty cuts presented in \S~\ref{subsec:observations}. The Halo sample and the three GS/E samples increase in size by between 2--8~per~cent when these cuts are removed, but the changes to the results -- the values of the best-fitting parameters, the derived masses, and the overall narrative presented in \S~\ref{sec:results} -- are negligible. According to the BIC/AIC the best-fitting density profiles are the same for the Halo sample and each of the three GS/E samples, and the masses change by only a few per~cent (the fit to the AD sample increases by 0.01 dex), far smaller than the statistical uncertainty on each measurement. We also check the assumptions which go into the construction of the effective selection function and the density profile normalization factors, namely the choice of [Fe/H] and age ranges. While \citet{montalban21} show that typical GS/E stars ages are about 10~Gyr old, the distribution extends down to $\sim 7$~Gyr old. Additionally, with regards to our choice of [Fe/H] range, whereas we elected to use a minimum [Fe/H] of -3 for the halo sample, the data only barely extend lower than [Fe/H] of -2. We replicate our fits but using a new isochrone grid which extends from 7-14~Gyr old (therefore rebuilding our effective selection function) and using a minimum [Fe/H] of -2 for the halo population (same as for the GS/E samples), but holding all other aspects of our analysis constant. We again find that this has a negligible impact on our results or overall narrative. Masses decrease by between 0.02 and 0.07 dex (the largest decrease is for the AD GS/E sample), best-fitting density profile parameters change little, and the AIC/BIC prefer the same density profiles. We therefore assign a systematic uncertainty which lowers the mass of GS/E by 0.07 dex (15~per~cent), or about $0.22\times10^{8}$~\Msun, due to choices in parameters for sample selection and effective selection function grid construction.

Second, as discussed when validating our method in \S~\ref{subsec:method-validation}, we expect the mass of our fits to be slightly overestimated due to contamination from the low-$\beta$ component. Now in the case of the mock data, both the low and high-$\beta$ components were drawn from the same density profile, and therefore one would only expect bias due to the excess in number counts (modulo some higher order effects from the spatial dependence of the selection). Secondarily, and as discussed above in the context of the disparity between \eLz\ and AD fits versus \JRLz\ and whole-halo fits, when the underlying density profiles differ then contamination from the low-$\beta$ component can lead to a best-fitting density profile which more closely resembles that of the low-$\beta$ component than the target density profile of the high-$\beta$ component. Given that we assume the low-$\beta$ component has a `more massive' density profile (i.e. the whole halo has a steeper power law in the inner Galaxy and a less marked break at intermediate Galactocentric radii) then we might also expect that contamination leads to an overestimation of the mass.

The first of these effects is simple to gauge: we expect to overestimate the mass by a factor equal to the inverse of the purity of the selection (this is neatly observed when determining the mass of our mocks). In the case of the AD selection the purity is estimated to be 0.9 for our initial kinematic effective selection function, and drops to 0.78 for our more realistic kinematic effective selection function created in \S~\ref{subsec:altering-kinematic-effective-selection-function}. The purity found by \citetalias{lane22} for the AD selection was 0.86. The second effect, the biasing of the shape of the density profile itself is challenging to gauge. We have posited above that contamination from the low-$\beta$ halo causes the increased mass found with the \JRLz\ fit, however this is higher than that for the AD fits by a factor of 3-4. Without any good way to gauge the impact of the contamination on the shape of the GS/E halo density profile we therefore fall back on the simple, conservative estimate being of order the inverse purity. We therefore adopt as our first systematic uncertainty one which lowers the mass to 0.78 (the lowest purity estimate for the AD selection) times the best-fitting mass, which is about $0.3\times10^{8}$~\Msun. We do note though that we have attempted to mitigate this source of systematic uncertainty by cutting stars with high [Al/Fe] from our kinematically-selected GS/E samples (see Figure~\ref{fig:halo_abundances}), since these stars are more likely to be part of the \textit{in-situ}, low-$\beta$ stellar halo.

Building on this, while we believe the trimming of stars with high [Al/Fe] is a well-motivated way to remove low-$\beta$ and thick disk contaminants from our kinematic selections, it is possible that these are genuine GS/E stars which should be included in our calculations. For the AD selection there are 4 stars which are removed on this basis, compared with 73 stars in the sample (there are similar fractions removed for the \eLz\ and \JRLz\ spaces, of order 5-8~per~cent). Assuming that inclusion of these stars would not change the density profile, only increase the normalization factor, this would result in an increase in the mass by $0.08\times10^{8}$~\Msun.

Finally, we consider that in \S~\ref{subsec:altering-kinematic-effective-selection-function} we investigated iterative improvements to the kinematic effective selection function, finding that more realistic kinematic effective selection functions constructed based on our fits generally have smaller values for a fixed kinematic selection. For example, referring to Figure~\ref{fig:ksf2_lb_dmod_marginalized} we see a typical value of 0.5 for the value of $\mathfrak{S}_{2}$ for the AD selection, while a typical value for the more realistic $\mathfrak{S}_{2}^{\prime}$ is more like 0.45. In \S~\ref{subsec:altering-kinematic-effective-selection-function} we specifically calculated a fractional decrease for the AD space of -0.07. Again, since this is for a fixed kinematic selection, and therefore a constant number of stars in the sample, we expect this decrease in the value of the kinematic effective selection function to manifest as an increase in the derived mass by an approximate factor $0.07$, or about $0.1\times10^{8}$~\Msun.

Before summarizing our systematics we discuss three related points. First, we must contend with the fact that we find a much lower fractional density for the high-$\beta$ GS/E component at the position of the Sun compared with other studies, which tend to attribute about 50~per~cent of the density near the Sun to GS/E \citep{belokurov18,lancaster19,iorio21}. We find that the density at the position of the Sun for the AD sample is $\nu_{0} = A\chi = 5.2^{+0.7}_{-0.6}\times 10^{3}$~\Msun~kpc$^{-3}$, compared with a density of $2.7^{+0.2}_{-0.2}\times 10^{4}$~\Msun~kpc$^{-3}$ for the halo sample. This is a difference by about a factor of 5, about 2.5 less than we would expect for an even mixture of high- and low-$\beta$ stars at the solar position. Since in our analysis mass scales directly with the density at the solar position, if we demand that the density of the GS/E component rises to the level of 50~per~cent of the whole-halo sample, then we would find the mass increases to about $3.8\times10^{8}$~\Msun. Now, we do not consider this to be a genuine source of systematic uncertainty, unlike those noted above, since we have no reason based on a sober assessment of our analysis to think that we have substantially underestimated the density at the solar position. For this reason we simply note that this is what our mass would become in light of these constraints from prior studies.

A second point is that we elected to model the high-$\beta$ portion of the stellar halo using an anisotropy of $\beta=0.8$ to avoid minor numerical instabilities noted by \citetalias{lane22} when using DFs with $\beta=0.9$. This is slightly lower than the value of $\beta \approx 0.9$ which is preferred by recent literature \citep{belokurov18,fattahi19,lancaster19,iorio21}. Our choice of $\beta=0.8$ actually implies that our derived mass is an upper limit, assuming that the geometry and form of the best-fitting GS/E density profile would be similar if we fit with a kinematic effective selection function derived using $\beta=0.9$. This is because the kinematic effective selection function should typically approach 1 as $\beta$ increases, which in turn decreases the derived mass. Intuitively, the selections from \citetalias{lane22} contain within them a larger fraction of a $\beta=0.9$ population than a $\beta=0.8$ population since they probe regions of kinematic space where extremely radial orbits lie, and therefore the required correction for stars missed by the restrictive kinematic cuts is smaller and the inferred mass is smaller. To check this explicitly, we compute a kinematic effective selection function assuming the same parameters as presented in \S~\ref{subsec:kinematic-effective-selection-function} except that we increase the high-$\beta$ anisotropy to 0.9. We find that the form of the correction is very similar to the $\beta=0.8$ correction both as a function of distance along each line of sight, as well as on-sky position, except that values are as expected systematically higher. For the AD space, we find typical values increase from $\sim 0.4$ to $\sim 0.6$, and similarly for the \eLz\ space we find typical values increase from $\sim 0.5$ to $\sim 0.7$. Assuming that were we to fit our samples with these kinematic effective selection functions that resulting density profiles would be similar to our current findings, we can say that the derived mass would be about 60-70~per~cent our current derived mass, or about $1\times10^{8}$~\Msun.

A third and final auxilliary point is to mention that while we are confident in our thick disk outlier model, it is possible that it does not fully capture all possible contaminants. It is well known that the thick disk has a range of geometries depending on the age and abundance of the specific sub-population \citep{bovy12,mackereth19b}, and the oldest, hottest, thickest parts may not be well-described by our simple single-exponential model. This is less likely to impact the measurement of the mass of GS/E, since these contaminants likely have higher [$\alpha$/Fe] and [Al/Fe] such that they are excluded from our chemical cuts on the GS/E subsamples. The whole-halo sample, on the other hand, may be plagued by such contaminants since it includes higher [Al/Fe] stars. It may be the case then, that our measurements of the total mass of the stellar halo are impacted, and by extension our estimates of the mass ratio of the GS/E merger. However since our stellar halo mass measurements are less than $10^{9}$~\Msun\ and therefore on the low end of modern estimates \citep[e.g.][]{deason19,mackereth20} it is likely that the degree of contamination is minimal.

Now by combining the sources of systematic uncertainty together, and adding uncertainties in quadrature, we arrive at an estimate for the mass including systematics of $1.45\ ^{+0.92}_{-0.51} \mathrm{(stat.)}\ ^{+0.13}_{-0.37} \mathrm{(sys.)}\ \times10^{8}$~\Msun. And to be clear we have not included the overestimations due to the discrepency between our measured fractional solar densities for GS/E and the literature estimates, or the modification of the mass due to a change in $\beta$ in this quoted value.

\subsection{Comparison of our findings with other work}
\label{subsec:comparison-with-other-work}

It is challenging to compare our results with those of other studies due to the unique nature by which we select our sample. Although the spirit of our kinematic and chemical selections is not that different from similar studies \citep[e.g.][]{mackereth20,han22}, our kinematic selections are far more restrictive and we correct for missing stars using a novel kinematic effective selection function. None the less, we attempt here to provide a context for the density profile parameters and mass of GS/E reported here.

\subsubsection{The mass of GS/E}

When it was first studied after its discovery, the stellar mass of the GS/E progenitor was thought to be quite high. \citet{helmi18} posited the stellar mass was $6\times10^{8}$~\Msun\ based on chemical abundance trends, and \citet{belokurov18} suggested that the virial mass of the progenitor was $> 10^{10}$~\Msun. Early simulation-based studies suggested that the stellar mass of the progenitor could by in the range $10^{8.5}-10^{10}$~\Msun\ \citep{fattahi19,mackereth19a}.  In particular, \citet{fattahi19} find that within 20~kpc the mass of the progenitor could be between about $2\times10^{8} - 2\times10^{9}$~\Msun. \citet{das20} estimate a total virial mass of about $10^{11}$~\Msun\ based on the dispersion of the remnant in action space, which equates to a stellar mass of about $10^{9.5}$~\Msun\ using canonical abundance matching relations. GS/E was also studied using chemistry, with \citet{vincenzo19} estimating a mass of $5\times10^{9}$\Msun\ and \citet{feuillet20} estimating a mass of $7\times10^{8} - 7\times10^{9}$~\Msun. \citet{naidu20} and \citet{naidu21} use H3 data -- and N-body simulations in the latter study -- to infer a mass in the range of $4-7\times10^{8}$~\Msun. The work of \citet{naidu21}, where the mass was found to be $5\times10^{8}$~\Msun, in particular is consistent with a number of auxilliary constraints on GS/E, such as the modestly retrograde tendancy for the debris \citep{belokurov18,helmi18}, the existence of a retrograde tail of debris at large $L_\mathrm{z}$ \citep{helmi18,naidu20}, as well as the concurrence and alignment with the major axes of the density ellipsoid of the debris with the Hercules-Aquila Cloud and the Virgo Overdensity \citep[first noted by][]{simion19}. More recently, \citet{rey23} use the VINTERGATEN suite of cosmological zoom simulations to study GS/E-like mergers of varying mass ratios using the `genetic modification' technique. Their results favour a similar mass ratio for the GS/E merger as our finding, but most interestingly they demonstrate that a wide range of merger mass ratios ($\sim 1:24 - 1:2$) can lead to very similar halo chemodynamics at $z=0$. This may help to explain why our results are at odds with many of the abundance-based measurements which predict a higher mass for GS/E.

The work of \citetalias{mackereth20} is the most closely analogous to ours, since they use a very similar technique, similar density profiles, and similar sample (they use APOGEE DR14). They find the mass of GS/E to be approximately $3\times10^{8}$~\Msun, although they base this estimate purely on a chemical selection of GS/E inspired by \citet{mackereth19a}. Another recent work which is very similar to ours is that of \citet{han22}, who fit triaxial broken power law density profiles to H3 data, finding a mass of $5.8-7.6\times10^{8}$~\Msun, with the lower mass end of that range corresponding to more restrictive eccentricity cuts on the GS/E sample. Compared to all of these works, we find a much lower mass, with a value of $1.45\times10^{8}$~\Msun\ derived for our most favoured kinematic selection and density profile. While it does bring itself into closer alignment with other findings, we no not favour the results of our \JRLz\ sample due to contamination issues as discussed in previous sections.

Another set of constraints on the mass of GS/E comes from the study of globular clusters systems associated with the remnant \citep[as tabulated by e.g.][]{myeong18,massari19}, for which the age-metallicity relation may be linked to the total mass of the host. \citet{kruijssen20} use the E-MOSAICS simulations to train a neural network to predict the accretion redshift and mass of the major Milky Way accretion remnants. For GS/E they find a mass of $2.7\times 10^{8}$~\Msun\ with error bars that are large enough to be consistent with our preferred measurement. \citet{forbes20} link the total number of GS/E-attributed globular clusters to its halo mass, which can then be used to infer the stellar mass using canonical relations, which yields an estimate of approximately $8\times10^{8}$~\Msun . \citet{callingham22} perform a similar analysis, relying on simulations and kinematics for additional context when grouping GS/E globular clusters, and arrive at a mass for the remnant of about $3.2\times10^{8}$~\Msun. \citet{limberg22} carry out a joint analysis of stars and globular clusters associated with GS/E, additionally relying on the hypothesis that the cluster $\omega$ Centauri was a nuclear star cluster of GS/E, and estimate a stellar mass of $1.3\times 10^{9}$~\Msun\ for the progenitor system.

Many lines of evidence for the mass of GS/E rely on the stellar mass-metallicity relation. It is therefore useful to ask whether the mass we derive for GS/E is unphysical given that its mean [Fe/H] remains unchallenged \citep[cf. our Figure~\ref{fig:selection_abundances} with ][]{myeong19,monty20,hasselquist21,horta23a}. \citet{naidu22} collect the stellar mass and metallicity information for Milky Way stellar halo remnants from the H3 survey. Comparing with their figure~2, we see that were GS/E to have a mass $\sim 1.5\times10^{8}$~\Msun\ that it would still be consistent with the observed mass-[Fe/H] and mass-[$\alpha$/Fe] trends. In fact, GS/E would more closely resemble the Helmi streams \citep{helmi99} in both stellar mass and metallicity. Under the assumption that the masses of the progenitors of these two structures are similar, the mismatch in their [$\alpha$/Fe] abundances (of order 0.1 dex) may require explanation, although there is intrinsic scatter in the relation that may permit the deviation. It could perhaps be the case that the Helmi streams has lower [$\alpha$/Fe] because it experienced lower star formation efficiency for its mass \citep[e.g.][]{matsuno22}. \citet{mackereth19a} study the progenitors of high-eccentricity stellar halo debris in Milky Way analogs the EAGLE simulations in order to derive a correspondence between mass and metallicity for comparison with APOGEE. Their observed trends of progenitor mass with [Fe/H] and [$\alpha$/Fe] (see their figure~11) would by no means exclude a progenitor with stellar mass $\sim 1.5\times10^{8}$~\Msun\ given a typical [Fe/H] for GS/E of $-1.2$. \citet{horta23b} use the FIRE simulations to investigate the properties of merger remnants in the halos of Milky Way analogs, and their findings would indeed be challenging to reconcile with our low stellar mass for GS/E. They show that fully phase-mixed remnants with [Fe/H] comparable to GS/E tend to have higher stellar masses. These comparisons suggest that while there may be some tension in placing such a light GS/E in a cosmological context, it would be by no means an outlier when compared with other remnants in the Milky Way stellar halo.

When comparing our results both with all other mass constraints for GS/E, but also specifically those of \citet{han22} and \citet{mackereth20} who also study the GS/E stellar density profile directly, our mass is certainly among the lowest ever claimed. Our findings of a low mass require explanation when compared with these two studies since the techniques are so similar. We posit that the reason for this is that our kinematic selections are far more restrictive. \citetalias{lane22} compared literature kinematic selections for GS/E with their derived high-purity selections, finding that some commonly used selections for GS/E can achieve purity as low as 0.5-0.6, and many achieve purity in the range of 0.6-0.7. This adds a substantial degree of contamination to samples selected in this way, artificially increasing inferred masses. This effect is compounded, however, by the fact that a contaminated sample will also yield a different fit to the GS/E sample, one which is more likely to resemble the low-$\beta$ stellar halo.  We specifically attribute the large derived mass of the \JRLz\ selection sample to this exact behaviour, since the purity using our original DF models is 0.78 and it drops to 0.52 when using our improved kinematic effective selection function.

\subsubsection{The density profile of GS/E}

We find that the density profile of GS/E is described by a shallow power law in the inner parts of the Galaxy, and steepens between 20 and 30~kpc. These findings are in good agreement with recent studies such as \citet{han22}, who detect breaks at 12 and 28~kpc. Our power law models with two breaks find those radii to be at about 15-20~kpc and 30-40~kpc, which is broadly consistent with the findings of \citet{han22}. These authors find inner, middle, and outer power laws of 1.7, 3.1, and 4.6 respectively. These power laws are roughly consistent with both our BPL and DBPL findings for the AD selection, where for the BPL profile we have inner power laws of about 1.2 steepening to about 3.8, and for the DBPL profile the inner, middle, and outer power laws are about 0.9, 3.0, and 6.2. We are therefore shallower in the inner Galaxy and steeper in the outer Galaxy when compared with the results of \citet{han22}. Our break radius is also consistent with the findings of \citetalias{mackereth20} who see a break radius of 25~kpc. They do find a much steeper inner power law ($\alpha \sim 3.5$), but they were not specifically targeting GS/E in their fits. in general, the finding of a break radius between 15-30~kpc is in broad agreement with other literature findings \citep{sesar11,deason11,xue15,deason19}. While we do detect power law breaks at reasonable radii, the broken power law models with one or two breaks are never favoured compared with simpler models. It is likely that this is caused by the extreme paucity of stars in our GS/E samples and the relative lack of stars in our halo sample compared with, for example, the H3 dataset \citep{h3}. Upcoming surveys such as SDSS-V \citep{sdss5} will have vastly more stars with which to probe more complicated density profiles using techniques with restrictive samples such as we use here.

Interestingly, we find that our best-fitting density profiles for GS/E are similar in orientation to those found by \citet{iorio19}. Specifically, the major axis of our density profiles is rotated by $90\degr-100\degr$ (specifically $99.5^{+13.9}_{-14.9}$ for our best-fitting AD model) in the Galactocentric $X-Y$ plane towards the direction of Galactic rotation (positive $Y$), and inclined about $16\degr$ below the plane. \citet{iorio19} find that the principal axis is rotated about $20\degr$ from the Galactocentric Y-axis (about $70\degr$ towards anti-rotation from the X-axis) and and inclination of about $20\degr$ (oriented the same as our model). When we cast our rotation angles in the same frame as theirs we would find rotation of about $9.5\degr$ from the Y-axis and $80.5\degr$ towards anti-rotation from the X-axis. These results are very similar (cf. our Figure~\ref{fig:pcontour} with their figures~2 and 3), reinforcing the idea that the Hercules-Aquila Cloud and Virgo Overdensity are related to the GS/E remnant. Similar findings were also recently reported by \citet{han22}, although they find the rotation of the principal axis with respect to the Sun-Galactic centre axis to be only $24\degr$, but a similar inclination of the principal axis of $25\degr$.

Regarding axis scale ratios, we find that the GS/E remnant is significantly more triaxial than other studies have reported. \citet{han22} find axis ratios of 1:0.8:0.7 for the GS/E remnant, and \citet{iorio18} finds 1:0.8:0.5-0.6, while we find axis ratios 1:0.55:0.45. The cause of this mismatch could be attributed to our selections, and perhaps the GS/E is truly more triaxial than these other studies have noted. A more likely interpretation of the difference in these findings is that GS/E changes in shape with increasing Galactocentric radius. The model of \citet{iorio18} has variable flattening with radius, while the shape parameters of the models used by \citet{han22}, like ours, are fixed with radius. It may be that surveys which are sensitive to different ranges of Galactocentric radius (due to the differences in observational footprint, choice of tracer, etc.) find different shape parameters for GS/E. It is curious though that the results of \citet{han22}, using data from the H3 survey which should probe larger Galactocentric radii compared with APOGEE which probe smaller Galactocentric radii, should find less triaxiality since one might generally expect large degrees of triaxiality at larger Galactocentric radii. Nevertheless, when these results are collated with recent findings of complex structure associated with the GS/E remnant at large Galactocentric radii by \citet{chandra23}, we can likely say that GS/E has a complicated shape-radius dependency.


\section{Summary and Conclusions}
\label{sec:summary-conclusions}

In this work we have measured the mass of the GS/E remnant using high purity samples of APOGEE DR16 red giants selected using kinematics and chemistry. We employ a density modelling approach which accounts for the APOGEE survey selection function, dust obscuration, and the density of the tracer population in colour-magnitude space. To this method we add a novel technique for correcting the biases induced by kinematic selections which is based the creation of a kinematic effective selection function using fiducial distribution function models of the Milky Way stellar halo. We identify GS/E in multiple kinematic spaces using high-purity kinematic selections based on the work of \citetalias{lane22}, and we fit triaxial rotated density profiles of varying radial complexity to these samples. We validate our density modelling approach, and specifically our new kinematic effective selection functions, using mock APOGEE data. Our main results can be summarized as follows:

\begin{itemize}
    \item We find that our kinematically-selected GS/E samples are well fit by density profiles which have relatively shallow or flat power law indices in the inner Galaxy ($0.5 < \alpha_{1} < 1.5$) and drop sharply at $\sim~20-30$~kpc to $\alpha>3.5$ at larger radii.
    
    \item The best-fitting GS/E density profiles are triaxial, with axis ratios $1:0.55:0.45$. The principal axis is rotated by about $100\degr$ towards the direction of Galactic rotation and is inclined $16\degr$ below the plane, which is in good agreement with previous studies.

    \item We determine the mass of GS/E to be $1.45\ ^{+0.92}_{-0.51}\,10^8\,M_\odot$, which is lower than found in previous studies. An assessment of sources of systematic error suggests they are smaller in magnitude than the statistical uncertainty quoted above. We attribute our finding of a lower mass to our use of restrictive kinematic cuts which produces a high-purity sample of GS/E stars and removes contamination from stellar halo populations with lower velocity anisotropy.

    \item A stellar mass of $1.5\times10^8\,\Msun$ corresponds to a total halo mass of GS/E of $\approx 5-6\times10^{10}\,\Msun$ using standard stellar-mass--to--halo-mass relations. Given that the Milky Way's halo mass at the time of the merger was likely $\approx 5\times 10^{11}\,\Msun$, GS/E constituted a minor 1:8-1:10 merger.

    \item We also fit density profiles to the whole stellar halo, finding that the best-fitting profiles have power law index around $\alpha=2.5$, with no evidence for breaks in the density profile. We find the mass of the stellar halo to be between $6.7-8.4\times10^{8}$~\Msun\ out to 70~kpc. GS/E therefore represents only a modest fraction of the mass of the stellar halo, about 15~per~cent at the low end or 25~per~cent at the high end.
\end{itemize}

We have shown evidence that the GS/E remnant is lighter than many previous studies have suggested. We argue that our precise kinematic selections for GS/E driven by fiducial stellar halo distribution function models leads to less contamination by stellar halo populations with low velocity anisotropy. While we are confident in our results, they represent the first application of a new technique for integrating kinematic selections into Galactic density modelling through the use of distribution functions. Additional work needs to be done to explore distribution functions which more accurately represent the stellar populations known to exist in the stellar halo, including triaxial, rotating distribution functions for example. We hope that simulations may be illuminating for this purpose, and plan to study this technique in the context of cosmological N-body simulations. We envision a future where stellar halo density modelling does not rely on kinematic selections to define samples, but involves modelling of the full six dimensions of phase space.

\section*{Acknowledgements}

We first thank the referee for their comments, which have certainly improved the quality of the manuscript. JMML and JB acknowledge financial support from NSERC (funding reference number RGPIN-2020-04712) and an Ontario Early Researcher Award (ER16-12-061). We thank Josh Speagle and Ting Li for helpful discussions and comments on the manuscript. This work has made use of data from the European Space Agency (ESA) mission {\it Gaia} (\url{https://www.cosmos.esa.int/gaia}), processed by the {\it Gaia} Data Processing and Analysis Consortium (DPAC, \url{https://www.cosmos.esa.int/web/gaia/dpac/consortium}). Funding for the DPAC has been provided by national institutions, in particular the institutions participating in the {\it Gaia} Multilateral Agreement. Funding for the Sloan Digital Sky Survey IV has been provided by the Alfred P. Sloan Foundation, the U.S. Department of Energy Office of Science, and the Participating Institutions. SDSS-IV acknowledges support and resources from the Center for High-Performance Computing at the University of Utah. The SDSS web site is \url{www.sdss.org}.

\section*{Data Availability}

The APOGEE DR16 data used in this article are available at: \url{https://www.sdss.org/dr16}. The \textit{Gaia} data used in this article are available at: \url{https://gea.esac.esa.int/archive/}.



\bibliographystyle{mnras}
\bibliography{manuscript}



\appendix

\section{Validation of the Method with Mock Data}
\label{ap:mock-data}

\begin{figure*}
    \centering
    \includegraphics[width=\linewidth]{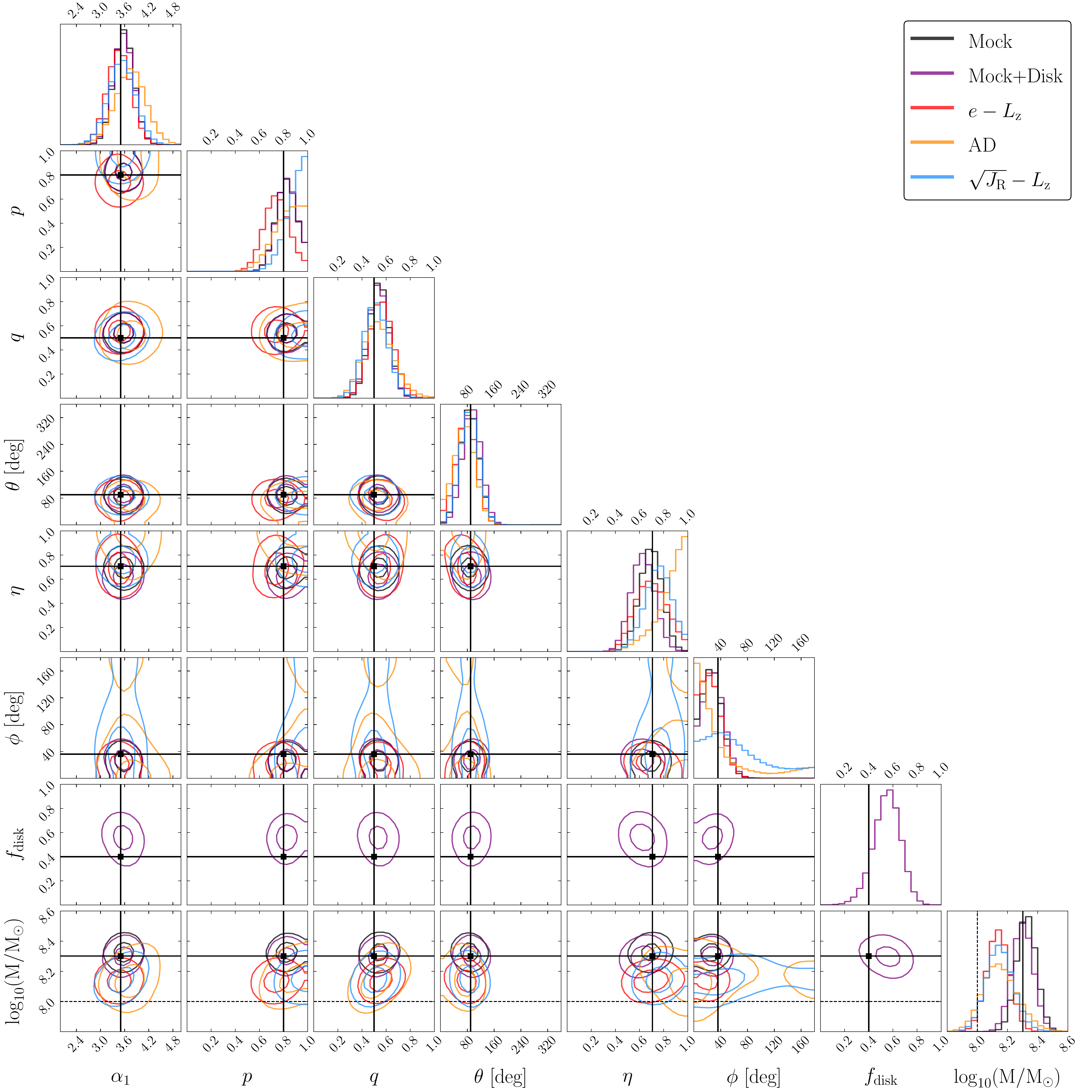}
    \caption{Corner plot showing the posterior samples of the fits to mock data. Contours are placed at the 1 and 2-$\sigma$ levels. The black lines show the fit to the whole mock sample without disk contamination, the grey line shows the fit including to the whole sample including disk contamination, and the coloured lines show the fits to the mock GS/E kinematic samples without disk contamination. the solid black lines show the input parameters for the models, and the dashed black lines show the mass expected for the kinematically selected GS/E analog populations (half that of the whole mock).}
    \label{fig:mock_posterior}
\end{figure*}

Here we show the results of applying our density fitting technique to mock data, as described in \S~\ref{subsec:method-validation}. Figure~\ref{fig:mock_posterior} shows the posterior distributions for fits to mock data with and without disk contamination, as well as mock data created using kinematic cuts that mirror those used to select the GS/E samples. The contours of the posterior distributions are set at 1 and 2-$\sigma$. The black lines show the input parameters for the models. The posterior distributions always contain the input model parameters, yet are broad in general. This is due to the fact that we chose to fit a lightweight mock halo (only $2\times10^{8}$~M$_{\odot}$ total mass) to test whether our modelling approach would work with a small number of stars. Since each stellar halo realization is subject to a certain degree of randomness we generate multiple stellar halos with similar parameters and a variety of masses. We find that the input parameters are always contained in the posterior, and that there are no major systematic over- or underestimations of input parameters beyond those noted in \S~\ref{subsec:method-validation}.

\section{Completeness of the modified kinematic effective selection function}
\label{ap:modified-ksf}

\begin{figure*}
    \centering
    \includegraphics[width=\textwidth]{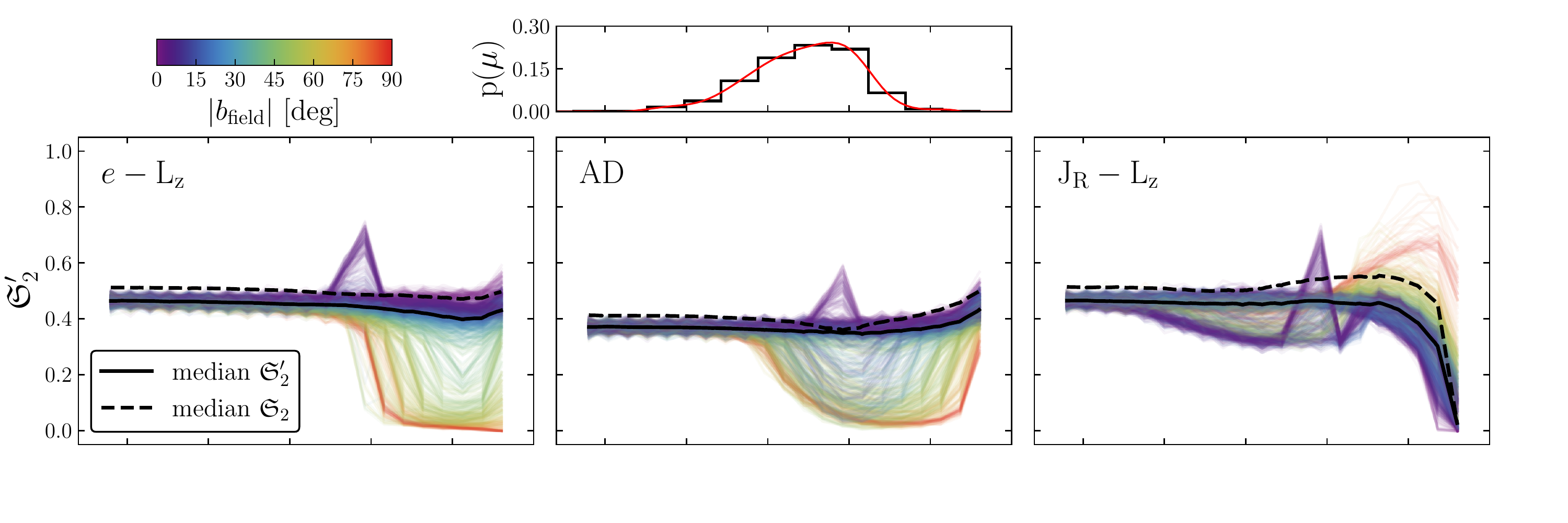}
    \caption{Same as Figure~\ref{fig:ksf_dmod_fields} but for the secondary kinematic effective selection function (denoted $\mathfrak{S}_{2}^{\prime}$) described in \S~\ref{subsec:altering-kinematic-effective-selection-function}. Overlaid on each panel are the median, calculated at each distane modulus, both of this secondary kinematic effective selection function $\mathfrak{S}_{2}^{\prime}$ (black solid line) and the original kinematic effective selection function $\mathfrak{S}_{2}$ (black dashed line) presented in Figure~\ref{fig:ksf2_dmod_fields}.}
    \label{fig:ksf2_dmod_fields}
\end{figure*}

\begin{figure}
    \centering
    \includegraphics[width=\linewidth]{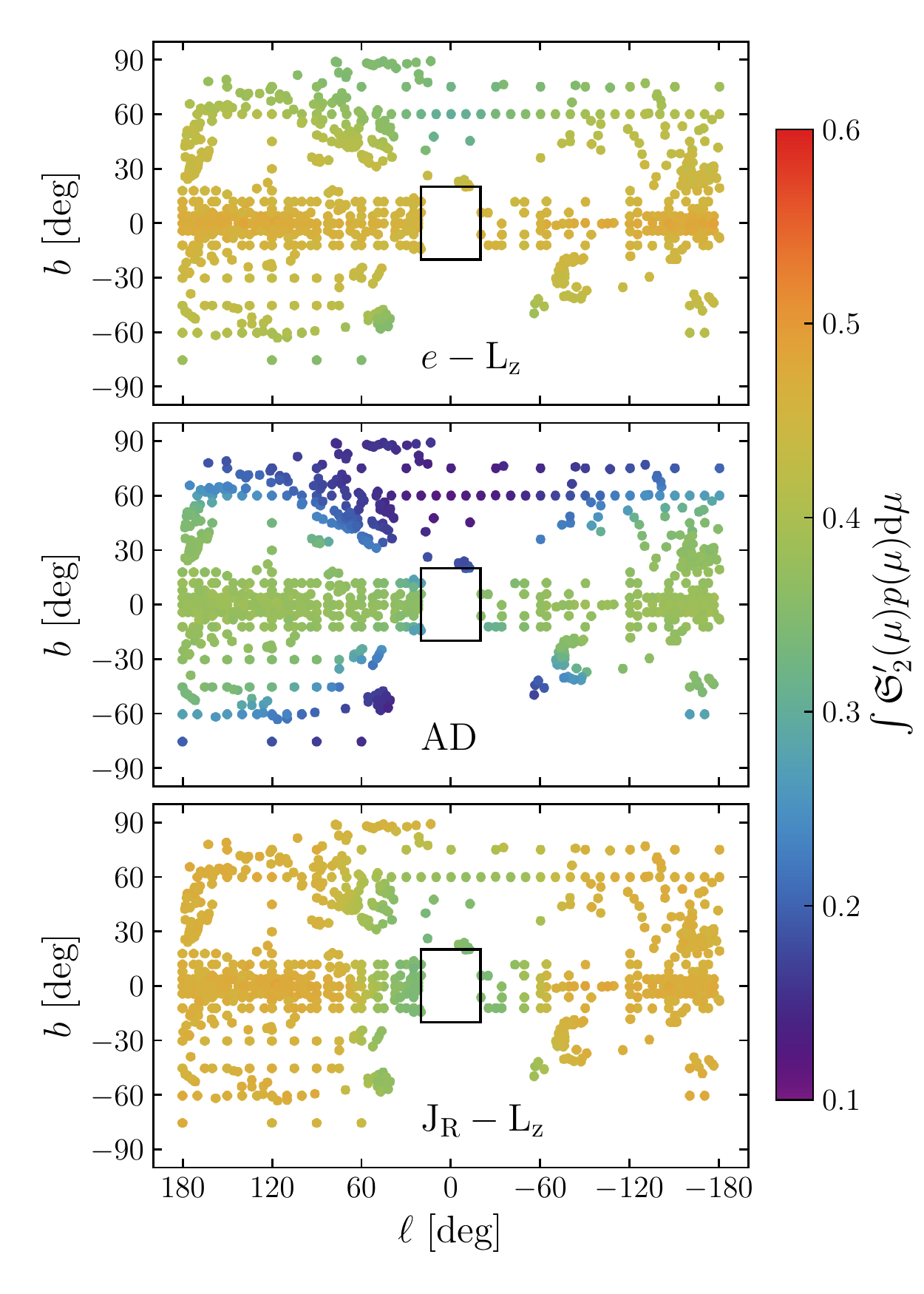}
    \caption{Same as Figure~\ref{fig:ksf_lb_dmod_marginalized} but for the secondary kinematic effective selection function described in \S~\ref{subsec:altering-kinematic-effective-selection-function}.}
    \label{fig:ksf2_lb_dmod_marginalized}
\end{figure}

Here we show the kinematic effective selection function derived based on our fits to the GS/E and whole-halo populations described in \S~\ref{subsec:altering-kinematic-effective-selection-function}. Figures~\ref{fig:ksf2_dmod_fields} and \ref{fig:ksf2_lb_dmod_marginalized} are identical to Figures~\ref{fig:ksf_dmod_fields} and \ref{fig:ksf_lb_dmod_marginalized}, except they use this new kinematic effective selection function (denoted $\mathfrak{S}_{2}^{\prime}$). On to Figure~\ref{fig:ksf2_dmod_fields} we add the median of $\mathfrak{S}_{2}^{\prime}$ and $\mathfrak{S}_{2}$, calculated across all fields at each distance modulus, to aid efficient comparison of typical values.

\bsp	
\label{lastpage}
\end{document}

%% file: tab-structural.tex
\begingroup
\setlength{\tabcolsep}{4pt} 
\renewcommand{\arraystretch}{1.} 
\begin{table*}
\centering
\setlength{\extrarowheight}{5pt}
\caption{Results of fits to the GS/E subsamples for each density profile and kinematic selection. The same results for the halo sample are also shown. Profiles fit are: ``SPL'': single power law; ``SC'': exponentially truncated single
power law; ``BPL'': broken power law; ``DBPL'': double broken power law. The suffix ``+D'' indicates that the fit includes a model for disk contamination. The models in each section with grey fill are those chosen to be best-fits based on assessment of the likelihoods presented in Table~\ref{tab:likelihood}.}
\label{tab:structural}

\resizebox{\textwidth}{!}{
\begin{tabular}{lcccccccccccc}

\hline 
Name & $\alpha_{1}$ & $\alpha_{2}$ & $\alpha_{3}$ & $r_{1}$ & $r_{2}$ & $p$ & $q$ & $\theta$ & $\eta$ & $\phi$ & $f_\mathrm{disk}$ & $\log_{10}(\mathrm{M}/\mathrm{M_{\odot}})$ \\ 
 &  &  &  & $[\mathrm{kpc}]$ & $[\mathrm{kpc}]$ &  &  & $[\mathrm{deg}]$ &  & $[\mathrm{deg}]$ &  &  \\ 
\hline 
\multicolumn{13}{c}{$e-L_\mathrm{z}$} \\ 
\hline 
SPL & $2.64^{+0.21}_{-0.21}$ & $-$ & $-$ & $-$ & $-$ & $0.91^{+0.07}_{-0.1}$ & $0.44^{+0.07}_{-0.06}$ & $122.9^{+177.0}_{-64.6}$ & $0.99^{+0.0}_{-0.01}$ & $94.9^{+59.6}_{-66.5}$ & $-$ & $8.2^{+0.07}_{-0.06}$ \\ 
SPL+D & $2.53^{+0.23}_{-0.24}$ & $-$ & $-$ & $-$ & $-$ & $0.89^{+0.07}_{-0.11}$ & $0.48^{+0.11}_{-0.08}$ & $110.0^{+187.6}_{-59.0}$ & $0.99^{+0.01}_{-0.03}$ & $90.9^{+59.1}_{-60.5}$ & $0.38^{+0.23}_{-0.24}$ & $8.21^{+0.08}_{-0.07}$ \\ 
SC & $-1.34^{+1.26}_{-1.29}$ & $-$ & $-$ & $4.06^{+1.82}_{-1.06}$ & $-$ & $0.7^{+0.15}_{-0.14}$ & $0.39^{+0.08}_{-0.08}$ & $76.4^{+147.6}_{-42.3}$ & $0.99^{+0.01}_{-0.01}$ & $98.0^{+13.1}_{-15.1}$ & $-$ & $7.91^{+0.08}_{-0.06}$ \\ 
\rowcolor{lightgray} SC+D & $-3.22^{+1.75}_{-1.81}$ & $-$ & $-$ & $3.46^{+1.59}_{-0.92}$ & $-$ & $0.58^{+0.16}_{-0.15}$ & $0.36^{+0.1}_{-0.09}$ & $78.8^{+202.7}_{-48.6}$ & $0.99^{+0.01}_{-0.02}$ & $99.6^{+9.8}_{-11.3}$ & $0.55^{+0.14}_{-0.21}$ & $7.89^{+0.11}_{-0.08}$ \\ 
BPL & $1.13^{+0.51}_{-0.56}$ & $4.89^{+1.25}_{-0.98}$ & $-$ & $16.94^{+6.32}_{-3.82}$ & $-$ & $0.68^{+0.17}_{-0.17}$ & $0.38^{+0.1}_{-0.09}$ & $82.9^{+230.8}_{-53.9}$ & $0.99^{+0.01}_{-0.01}$ & $98.8^{+13.2}_{-13.6}$ & $-$ & $7.96^{+0.1}_{-0.08}$ \\ 
BPL+D & $0.71^{+0.54}_{-0.46}$ & $5.2^{+1.27}_{-1.05}$ & $-$ & $20.46^{+9.49}_{-5.62}$ & $-$ & $0.57^{+0.19}_{-0.18}$ & $0.35^{+0.12}_{-0.11}$ & $70.8^{+231.6}_{-44.1}$ & $0.99^{+0.01}_{-0.02}$ & $98.9^{+10.8}_{-10.5}$ & $0.56^{+0.15}_{-0.24}$ & $7.96^{+0.13}_{-0.1}$ \\ 
DBPL & $1.0^{+0.57}_{-0.59}$ & $4.03^{+1.22}_{-1.48}$ & $6.84^{+2.07}_{-1.83}$ & $15.14^{+5.24}_{-3.82}$ & $33.31^{+14.44}_{-11.93}$ & $0.69^{+0.16}_{-0.16}$ & $0.38^{+0.1}_{-0.09}$ & $84.7^{+230.1}_{-54.3}$ & $0.99^{+0.0}_{-0.01}$ & $98.8^{+13.5}_{-14.2}$ & $-$ & $7.94^{+0.09}_{-0.07}$ \\ 
DBPL+D & $0.64^{+0.58}_{-0.44}$ & $4.08^{+1.37}_{-2.09}$ & $6.8^{+2.11}_{-1.8}$ & $17.14^{+7.45}_{-4.82}$ & $34.17^{+13.65}_{-12.15}$ & $0.61^{+0.18}_{-0.18}$ & $0.37^{+0.12}_{-0.11}$ & $77.7^{+216.0}_{-48.9}$ & $0.99^{+0.01}_{-0.02}$ & $98.2^{+12.1}_{-12.1}$ & $0.55^{+0.15}_{-0.24}$ & $7.92^{+0.11}_{-0.09}$ \\ 
\hline 
\multicolumn{13}{c}{AD} \\ 
\hline 
SPL & $2.17^{+0.3}_{-0.31}$ & $-$ & $-$ & $-$ & $-$ & $0.66^{+0.21}_{-0.2}$ & $0.61^{+0.24}_{-0.23}$ & $134.3^{+74.5}_{-85.2}$ & $0.8^{+0.15}_{-0.2}$ & $98.0^{+21.6}_{-20.6}$ & $-$ & $8.49^{+0.15}_{-0.14}$ \\ 
SPL+D & $2.06^{+0.32}_{-0.33}$ & $-$ & $-$ & $-$ & $-$ & $0.58^{+0.24}_{-0.2}$ & $0.56^{+0.26}_{-0.23}$ & $139.4^{+60.4}_{-96.4}$ & $0.77^{+0.16}_{-0.18}$ & $97.4^{+19.8}_{-17.3}$ & $0.47^{+0.23}_{-0.29}$ & $8.52^{+0.16}_{-0.14}$ \\ 
\rowcolor{lightgray} SC & $-0.57^{+1.58}_{-1.95}$ & $-$ & $-$ & $8.57^{+12.58}_{-4.36}$ & $-$ & $0.54^{+0.22}_{-0.2}$ & $0.46^{+0.27}_{-0.2}$ & $147.0^{+43.5}_{-95.3}$ & $0.84^{+0.12}_{-0.22}$ & $99.5^{+13.9}_{-14.9}$ & $-$ & $8.16^{+0.21}_{-0.19}$ \\ 
SC+D & $-0.22^{+1.36}_{-1.78}$ & $-$ & $-$ & $11.31^{+17.08}_{-6.14}$ & $-$ & $0.49^{+0.21}_{-0.19}$ & $0.46^{+0.31}_{-0.2}$ & $149.6^{+60.4}_{-106.0}$ & $0.8^{+0.14}_{-0.21}$ & $98.2^{+14.2}_{-13.8}$ & $0.37^{+0.25}_{-0.24}$ & $8.22^{+0.22}_{-0.21}$ \\ 
BPL & $1.05^{+0.86}_{-0.72}$ & $3.79^{+2.44}_{-0.98}$ & $-$ & $23.59^{+15.55}_{-9.3}$ & $-$ & $0.55^{+0.22}_{-0.2}$ & $0.47^{+0.28}_{-0.2}$ & $146.6^{+48.8}_{-97.7}$ & $0.84^{+0.12}_{-0.22}$ & $98.5^{+15.0}_{-15.4}$ & $-$ & $8.24^{+0.18}_{-0.18}$ \\ 
BPL+D & $1.21^{+0.7}_{-0.8}$ & $3.91^{+3.1}_{-1.18}$ & $-$ & $26.95^{+15.62}_{-11.57}$ & $-$ & $0.51^{+0.24}_{-0.18}$ & $0.48^{+0.29}_{-0.21}$ & $151.8^{+54.3}_{-108.8}$ & $0.79^{+0.15}_{-0.2}$ & $98.2^{+15.7}_{-15.5}$ & $0.39^{+0.25}_{-0.26}$ & $8.25^{+0.21}_{-0.19}$ \\ 
DBPL & $0.85^{+0.81}_{-0.6}$ & $3.0^{+1.27}_{-0.98}$ & $6.15^{+2.62}_{-2.26}$ & $17.6^{+10.31}_{-5.81}$ & $39.5^{+10.59}_{-12.0}$ & $0.57^{+0.21}_{-0.18}$ & $0.49^{+0.25}_{-0.2}$ & $149.1^{+41.9}_{-99.8}$ & $0.84^{+0.12}_{-0.22}$ & $99.6^{+15.5}_{-15.9}$ & $-$ & $8.15^{+0.17}_{-0.16}$ \\ 
DBPL+D & $0.88^{+0.79}_{-0.63}$ & $2.9^{+1.46}_{-0.95}$ & $6.6^{+2.34}_{-2.6}$ & $18.62^{+10.93}_{-6.7}$ & $40.78^{+9.87}_{-12.08}$ & $0.54^{+0.22}_{-0.16}$ & $0.5^{+0.27}_{-0.2}$ & $151.6^{+50.9}_{-109.3}$ & $0.81^{+0.14}_{-0.21}$ & $98.3^{+15.7}_{-16.0}$ & $0.38^{+0.23}_{-0.24}$ & $8.14^{+0.18}_{-0.16}$ \\ 
\hline 
\multicolumn{13}{c}{$\sqrt{J_\mathrm{R}}-L_\mathrm{z}$} \\ 
\hline 
SPL & $2.55^{+0.18}_{-0.18}$ & $-$ & $-$ & $-$ & $-$ & $0.62^{+0.08}_{-0.08}$ & $0.67^{+0.12}_{-0.1}$ & $170.9^{+52.7}_{-146.5}$ & $0.87^{+0.09}_{-0.22}$ & $105.2^{+7.5}_{-7.8}$ & $-$ & $8.75^{+0.05}_{-0.05}$ \\ 
\rowcolor{lightgray} SPL+D & $2.39^{+0.2}_{-0.2}$ & $-$ & $-$ & $-$ & $-$ & $0.57^{+0.08}_{-0.08}$ & $0.69^{+0.15}_{-0.12}$ & $167.7^{+90.4}_{-94.8}$ & $0.93^{+0.05}_{-0.16}$ & $105.4^{+6.8}_{-6.8}$ & $0.63^{+0.11}_{-0.17}$ & $8.78^{+0.07}_{-0.06}$ \\ 
SC & $2.02^{+0.26}_{-0.33}$ & $-$ & $-$ & $37.24^{+12.13}_{-14.3}$ & $-$ & $0.61^{+0.08}_{-0.08}$ & $0.67^{+0.13}_{-0.1}$ & $171.0^{+58.6}_{-143.6}$ & $0.89^{+0.08}_{-0.23}$ & $106.1^{+7.0}_{-7.6}$ & $-$ & $8.64^{+0.05}_{-0.05}$ \\ 
SC+D & $1.7^{+0.33}_{-0.5}$ & $-$ & $-$ & $33.62^{+14.26}_{-14.04}$ & $-$ & $0.54^{+0.09}_{-0.09}$ & $0.7^{+0.16}_{-0.14}$ & $167.8^{+98.7}_{-71.8}$ & $0.94^{+0.05}_{-0.15}$ & $106.7^{+6.5}_{-6.7}$ & $0.66^{+0.1}_{-0.16}$ & $8.66^{+0.07}_{-0.07}$ \\ 
BPL & $2.48^{+0.22}_{-0.54}$ & $3.81^{+3.52}_{-1.15}$ & $-$ & $39.93^{+11.21}_{-29.1}$ & $-$ & $0.62^{+0.08}_{-0.08}$ & $0.69^{+0.13}_{-0.11}$ & $174.9^{+56.5}_{-142.2}$ & $0.88^{+0.09}_{-0.2}$ & $105.9^{+7.4}_{-7.6}$ & $-$ & $8.67^{+0.07}_{-0.06}$ \\ 
BPL+D & $2.2^{+0.3}_{-1.06}$ & $3.16^{+3.27}_{-0.72}$ & $-$ & $32.27^{+17.39}_{-21.93}$ & $-$ & $0.56^{+0.09}_{-0.09}$ & $0.71^{+0.15}_{-0.14}$ & $171.0^{+94.2}_{-76.0}$ & $0.94^{+0.05}_{-0.14}$ & $106.0^{+6.8}_{-6.9}$ & $0.64^{+0.1}_{-0.17}$ & $8.7^{+0.08}_{-0.08}$ \\ 
DBPL & $2.31^{+0.33}_{-1.32}$ & $2.82^{+1.39}_{-0.36}$ & $5.43^{+3.09}_{-2.23}$ & $17.59^{+22.88}_{-12.94}$ & $46.66^{+6.04}_{-13.84}$ & $0.63^{+0.08}_{-0.07}$ & $0.71^{+0.13}_{-0.1}$ & $175.7^{+61.1}_{-132.2}$ & $0.89^{+0.08}_{-0.2}$ & $106.3^{+7.3}_{-7.7}$ & $-$ & $8.63^{+0.06}_{-0.06}$ \\ 
DBPL+D & $1.89^{+0.51}_{-1.21}$ & $2.62^{+1.14}_{-0.37}$ & $5.04^{+3.16}_{-1.99}$ & $14.89^{+21.48}_{-8.92}$ & $46.1^{+6.5}_{-13.64}$ & $0.57^{+0.08}_{-0.08}$ & $0.74^{+0.15}_{-0.14}$ & $175.7^{+94.3}_{-67.3}$ & $0.93^{+0.05}_{-0.13}$ & $107.3^{+6.8}_{-7.0}$ & $0.64^{+0.1}_{-0.15}$ & $8.64^{+0.08}_{-0.08}$ \\ 
\hline 
\multicolumn{13}{c}{Halo} \\ 
\hline 
SPL & $2.84^{+0.07}_{-0.07}$ & $-$ & $-$ & $-$ & $-$ & $0.9^{+0.05}_{-0.05}$ & $0.55^{+0.03}_{-0.03}$ & $270.2^{+51.2}_{-202.9}$ & $1.0^{+0.0}_{-0.0}$ & $94.7^{+12.8}_{-13.1}$ & $-$ & $8.9^{+0.02}_{-0.02}$ \\ 
\rowcolor{lightgray} SPL+D & $2.49^{+0.09}_{-0.09}$ & $-$ & $-$ & $-$ & $-$ & $0.79^{+0.06}_{-0.06}$ & $0.58^{+0.05}_{-0.05}$ & $241.2^{+58.2}_{-89.2}$ & $1.0^{+0.0}_{-0.01}$ & $99.7^{+7.5}_{-7.5}$ & $0.73^{+0.04}_{-0.04}$ & $8.94^{+0.03}_{-0.03}$ \\ 
SC & $2.59^{+0.08}_{-0.08}$ & $-$ & $-$ & $49.65^{+3.93}_{-6.9}$ & $-$ & $0.89^{+0.05}_{-0.04}$ & $0.56^{+0.03}_{-0.03}$ & $272.8^{+52.2}_{-220.9}$ & $1.0^{+0.0}_{-0.0}$ & $96.0^{+11.0}_{-11.2}$ & $-$ & $8.83^{+0.02}_{-0.02}$ \\ 
SC+D & $2.14^{+0.12}_{-0.14}$ & $-$ & $-$ & $45.38^{+6.87}_{-10.24}$ & $-$ & $0.76^{+0.06}_{-0.06}$ & $0.59^{+0.05}_{-0.05}$ & $244.6^{+62.2}_{-99.2}$ & $0.99^{+0.0}_{-0.01}$ & $100.7^{+6.8}_{-7.0}$ & $0.75^{+0.03}_{-0.04}$ & $8.84^{+0.03}_{-0.03}$ \\ 
BPL & $2.84^{+0.07}_{-0.07}$ & $4.87^{+3.0}_{-1.68}$ & $-$ & $48.64^{+4.54}_{-7.54}$ & $-$ & $0.91^{+0.05}_{-0.05}$ & $0.56^{+0.03}_{-0.03}$ & $275.1^{+47.3}_{-214.7}$ & $1.0^{+0.0}_{-0.0}$ & $94.6^{+14.4}_{-14.0}$ & $-$ & $8.86^{+0.03}_{-0.03}$ \\ 
BPL+D & $2.4^{+0.15}_{-1.21}$ & $2.73^{+3.41}_{-0.24}$ & $-$ & $32.17^{+19.1}_{-27.56}$ & $-$ & $0.79^{+0.06}_{-0.06}$ & $0.59^{+0.05}_{-0.05}$ & $245.9^{+57.9}_{-97.9}$ & $1.0^{+0.0}_{-0.01}$ & $100.0^{+7.6}_{-7.6}$ & $0.73^{+0.04}_{-0.05}$ & $8.9^{+0.04}_{-0.06}$ \\ 
DBPL & $1.03^{+1.12}_{-0.73}$ & $2.77^{+0.12}_{-1.13}$ & $2.95^{+1.98}_{-0.12}$ & $3.94^{+0.76}_{-1.25}$ & $10.9^{+39.26}_{-6.49}$ & $0.89^{+0.05}_{-0.05}$ & $0.54^{+0.03}_{-0.03}$ & $271.2^{+48.3}_{-200.4}$ & $1.0^{+0.0}_{-0.0}$ & $95.4^{+11.5}_{-11.6}$ & $-$ & $8.87^{+0.03}_{-0.03}$ \\ 
DBPL+D & $1.71^{+0.74}_{-1.18}$ & $2.53^{+0.21}_{-0.24}$ & $3.67^{+3.82}_{-1.1}$ & $5.34^{+24.55}_{-1.97}$ & $45.84^{+6.82}_{-36.13}$ & $0.78^{+0.06}_{-0.06}$ & $0.59^{+0.05}_{-0.05}$ & $252.0^{+55.9}_{-94.4}$ & $0.99^{+0.0}_{-0.01}$ & $100.7^{+7.2}_{-7.7}$ & $0.74^{+0.04}_{-0.04}$ & $8.87^{+0.06}_{-0.05}$ \\ 
\hline

\end{tabular}
}

\end{table*}
\endgroup

%% file: tab-likelihood.tex
\begin{table}
\centering
\small
\caption{Likelihoods, AIC, and BIC for the fits to each of the GS/E subsamples for each of the density profiles. The models in each section with grey fill are those chosen to be the best-fits based on the maximum likelihoods, AIC, and BIC values.}
\label{tab:likelihood}
\begin{tabular}{lccc}

\hline 
Name & $\mathrm{max}(\mathcal{L})$ & AIC & BIC \\ 
\hline 
\multicolumn{4}{c}{$e-L_\mathrm{z}$} \\ 
\hline 
SPL & 0.0 & 0.0 & 0.0 \\ 
SPL+D & -0.0 & 2.1 & 5.1 \\ 
SC & 10.7 & -19.3 & -16.3 \\ 
\rowcolor{lightgray} SC+D & 13.1 & -22.2 & -16.1 \\ 
BPL & 10.6 & -17.2 & -11.1 \\ 
BPL+D & 13.1 & -20.2 & -11.1 \\ 
DBPL & 11.8 & -15.7 & -3.6 \\ 
DBPL+D & 13.6 & -17.3 & -2.2 \\ 
\hline 
\multicolumn{4}{c}{AD} \\ 
\hline 
SPL & 0.0 & 0.0 & 0.0 \\ 
SPL+D & 0.5 & 1.0 & 3.3 \\ 
\rowcolor{lightgray} SC & 5.7 & -9.4 & -7.1 \\ 
SC+D & 5.8 & -7.5 & -2.9 \\ 
BPL & 5.8 & -7.6 & -3.0 \\ 
BPL+D & 5.8 & -5.7 & 1.2 \\ 
DBPL & 6.4 & -4.9 & 4.3 \\ 
DBPL+D & 6.4 & -2.8 & 8.7 \\ 
\hline 
\multicolumn{4}{c}{$\sqrt{J_\mathrm{R}}-L_\mathrm{z}$} \\ 
\hline 
SPL & 0.0 & 0.0 & 0.0 \\ 
\rowcolor{lightgray} SPL+D & 4.3 & -6.5 & -2.8 \\ 
SC & 0.2 & 1.6 & 5.3 \\ 
SC+D & 4.9 & -5.9 & 1.6 \\ 
BPL & 0.9 & 2.2 & 9.6 \\ 
BPL+D & 6.1 & -6.3 & 4.8 \\ 
DBPL & 0.9 & 6.2 & 21.0 \\ 
DBPL+D & 6.1 & -2.3 & 16.3 \\ 
\hline 
\multicolumn{4}{c}{Halo} \\ 
\hline 
SPL & 0.0 & 0.0 & 0.0 \\ 
\rowcolor{lightgray} SPL+D & 27.2 & -52.3 & -47.0 \\ 
SC & -4.1 & 10.2 & 15.6 \\ 
SC+D & 26.1 & -48.3 & -37.6 \\ 
BPL & 2.7 & -1.3 & 9.3 \\ 
BPL+D & 28.3 & -50.5 & -34.6 \\ 
DBPL & 2.6 & 2.8 & 24.1 \\ 
DBPL+D & 28.7 & -47.3 & -20.7 \\ 
\hline

\end{tabular}
\end{table}


